# Giant optical spin-orbit interactions in ferroelectric van der Waals waveguides

Ding Xu[1†], Saeed Rahmanian Koshkaki[2†], Vicente Galicia[1], Chun-Ying Huang[1], Victoria Quirós-Cordero[1,3], Jakhangirkhodja A. Tulyagankhodjaev[1], André Koch Liston[1], Daniel G. Chica[1], Emma Lian[1], Amirhosein Amini[2], Yongseok Hong[1], Taketo Handa[1], P. James Schuck[3], Xiaoyang Zhu[1], Xavier Roy[1], Arkajit Mandal[2]*, Milan Delor[1]*

1. Department of Chemistry, Columbia University, New York, NY 10027, USA
2. Department of Chemistry, Texas A&M University, College Station, Texas 77843, USA
3. Department of Mechanical Engineering, Columbia University, New York, NY, 10027, USA

† Equal contribution

*mandal@tamu.edu; milan.delor@columbia.edu

**Abstract**

Optical spin-orbit interactions (SOI) link photonic spin to momentum, offering a route toward on-chip polarization control and beam steering. Nevertheless, achieving sufficient optical SOI and nonlinearities on sub-micrometer scales—a prerequisite for dense photonic integration—remains an outstanding challenge. Here, we show that highly birefringent van der Waals (vdW) waveguides provide an ideal, chip-compatible platform to address this limitation. We focus on the ferroelectric semiconductor $NbOI_2$, which exhibits record optical nonlinearities and dielectric anisotropy. Using femtosecond optical microscopy, we image light propagation and harmonic conversion beyond the total internal reflection barrier over tens of micrometers in $NbOI_2$ slab waveguides. We report giant optical spin-splitting through the optical spin Hall effect, which facilitates spatial separation of optical spin currents on sub-micrometer scales, in quantitative agreement with a microscopic light-matter interaction model. We further leverage optical spin-momentum locking to realize polarization-controlled waveguide steering. We generalize these observations across various vdW waveguides and empirically confirm a scaling law linking dielectric anisotropy to geometric spin-splitting. Our results establish highly anisotropic vdW waveguides as an ideal platform for densely integrated opto-spintronic technologies.

## Main

Spin-orbit interactions (SOI) in materials with broken symmetries underpin a range of emergent phenomena in condensed matter, including the spin Hall effect[1–3], topological phases[4], and unconventional superconductivity[5]. In analogy, confined light propagation in anisotropic media exhibits optical SOI driven by the momentum dependence of photon polarization eigenmodes[6,7]. This coupling between photonic spin (helicity) and momentum (wavevector) manifests as polarization-dependent routing and the optical spin Hall effect (OSHE)[6,8,9], offering a path toward on-chip polarization control and beam steering. Unlike electronic systems, photons maintain macroscopic coherence at room temperature, making photonic spin-momentum locking well suited for optical spin-based information encoding in practical devices. A key challenge for emerging optical technologies, such as optical computing, is to develop chip-compatible polarizers, beamsplitters, and nonlinear elements that function over short propagation distances. Materials with exceptionally strong optical SOI could serve as the foundation for these integrated optical architectures.

Progress in optimizing and leveraging optical SOI has been largely enabled by advances in material platforms. Early evidence of optical SOI dates back to the 1950's, when Fedorov predicted and later Imbert observed the polarization-dependent transverse displacement of light under total internal reflection.[13–15] The theoretical generalization of optical SOI under the framework of geometric phase[6,8,16] led to the first observations of the OSHE at dielectric interfaces[17,18] and in GaAs-based microcavities[9]. Recent tour-de-force experiments using high-quality halide perovskite microcavities, as well as liquid-crystal-tuned birefringent cavities, have realized pronounced OSHE and spin-polarized photon currents over extended propagation lengths[7,19–24]. However, current systems do not combine the SOI strength, chip compatibility, and optical nonlinearities required for light manipulation in densely integrated photonic technologies. Larger dielectric anisotropy combined with greater light confinement are required to reach this goal.

Here, we demonstrate that layered van der Waals (vdW) semiconductors are naturally suited for realizing strong optical SOI due to their large dielectric anisotropy,[25–29] strong light-matter interactions, large refractive indices, and layered geometries with atomically smooth interfaces enabling natural low-loss waveguides.[30,31] Among vdW materials, the ferroelectric niobium oxyhalides ($NbOX_2$) stand out for their giant nonlinearities arising from a second-order Jahn-Teller distortion coupled to polar order[10–12,32] and tunable optical response[33–35]. By directly imaging light transport in natural $NbOI_2$ slab waveguides, we show that the material's record dielectric anisotropy facilitates ultrafast spatial separation of polarized photon currents through double-refraction and the OSHE (Fig. 1a,b), as well as highly efficient polarization-controlled beam steering. These results are in quantitative agreement with quantum dynamics simulations from an anisotropic light-matter coupling model. We also show that spin-split photon currents generate second harmonic light, allowing us to simultaneously realize polarization control, beam steering, and nonlinear conversion in a chip-compatible waveguide. Our measurements are generalized across multiple vdW materials, revealing exceptionally large optical SOI in vdW waveguides compared to leading platforms, establishing a blueprint for densely integrated opto-spintronic technologies.

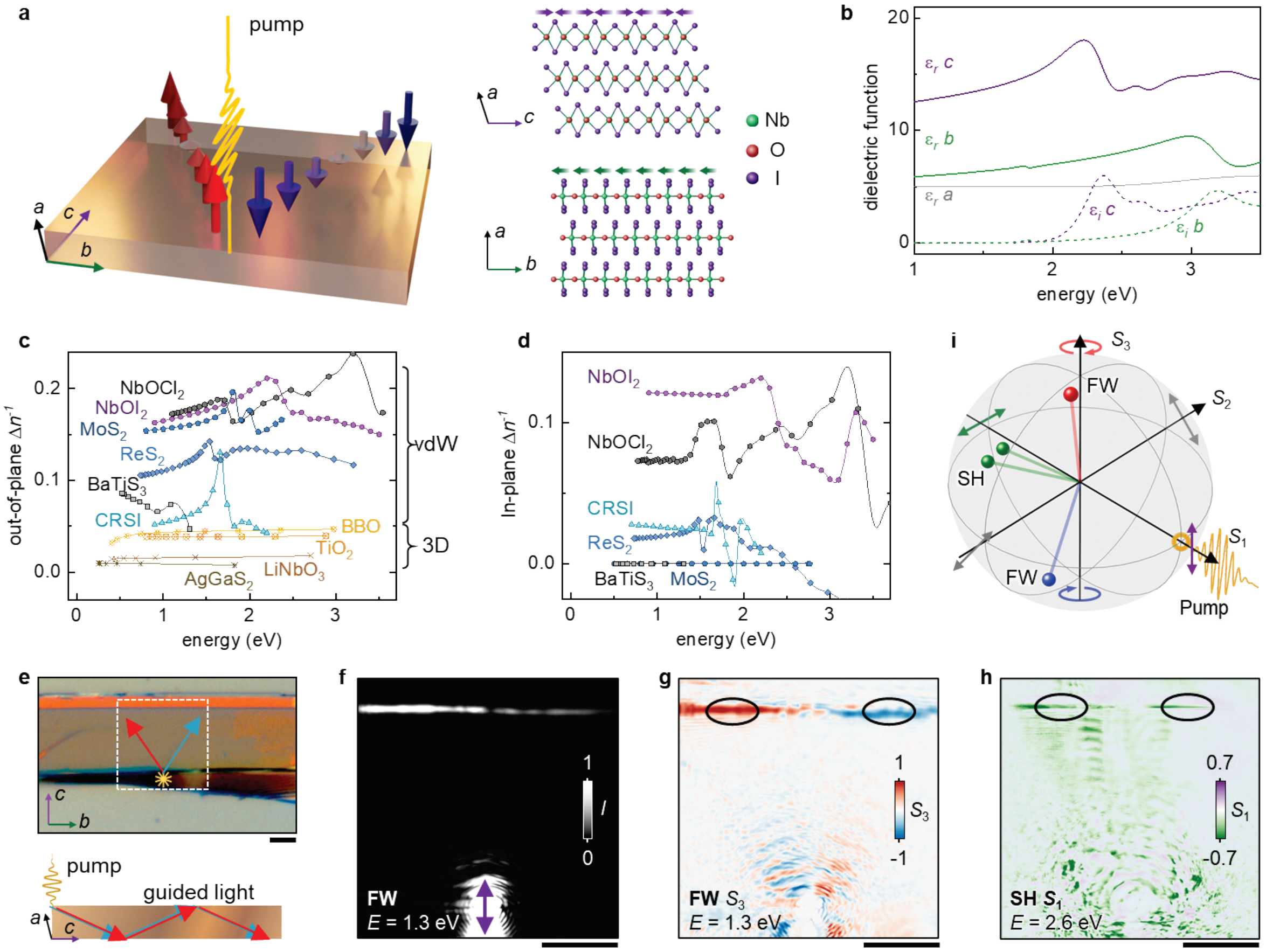

**Figure 1. Optical spin-splitting in $NbOI_2$ slab waveguides**. **a**, Schematic of optical spin splitting in a $NbOI_2$ planar waveguide. The incident light is focused on the $NbOI_2$ edge to couple into waveguide modes beyond the far-field light cone. Inset: side view of the $NbOI_2$ crystal structure along the non-polar *c*-axis and polar *b*-axis. **b**, Measured dielectric function of $NbOI_2$ along all crystal axes. Subscripts *r*, *i* indicate real and imaginary parts of the dielectric function, respectively. **c-d**, Comparison of out-of-plane and in-plane birefringence $\Delta n^{-1}$ for selected highly birefringent 3-dimensional and layered semiconductors and insulators. Dielectric functions other than for $NbOI_2$ and CRSI ($CsRe_6Se_8I_3$) are adopted from previously reported results.[26–29,36–40] **e**, Optical image of a 265 nm thick $NbOI_2$ planar waveguide on a glass substrate. The crystal is cleaved along the *b*-axis. The yellow star marks the pump excitation location. The red and blue arrows represent the spin-split beam trajectories that emerge from linearly polarized excitation of the edge. Inset shows a side-view schematic of the edge excitation. **f**, Far-field microscope image of waveguided light at the excitation energy of $E_{pump}$ = 1.3 eV, in the region highlighted by the dashed square in panel (e). The purple arrow indicates the pump polarization orientation. The image shows scattered light at the in-coupling and outcoupling edges of the waveguide. FW = fundamental wave. **g**, $S_3$ Stokes parameter imaging of the FW at 1.3 eV. **h**, $S_1$ Stokes parameter imaging of the second harmonic wave (SH), spectrally separated from the FW with a bandpass filter. **i**, Stokes vectors of the FW (blue and red) and SH (green) on the Poincaré sphere, measured at the output edge marked by circles in panels (g, h). All scale bars are 10 µm.

**Spin-split photon currents and nonlinear conversion in slab waveguides**

$NbOI_2$ is a monoclinic vdW semiconductor (space group *C2*)[10] that exhibits pronounced biaxial birefringence[41]. The Peierls-distorted, highly polarizable Nb-I chains along the *c*-axis contribute to a bandgap at 2.3 eV and a remarkably large dielectric constant ($\varepsilon_c \approx 12$) – similar to that of silicon despite the bandgap being more than twice as large. The more localized Nb-O bonding along the *b*-axis contribute to a wider bandgap (3.1 eV) and lower dielectric ($\varepsilon_b \approx 6$) (Fig. 1b). The alternating Nb-O bond lengths arising from a second-order Jahn-Teller distortion result in room-temperature ferroelectric order and giant second-order susceptibility $\chi^{(2)}$.[10,42] The highly anisotropic bonding configuration along *b*- and *c*-axes, as well as the layered structure, give rise to extraordinarily large dielectric anisotropy. Fig. 1c-d plots the inverse refractive-index difference, defined as $\Delta n^{-1} = |n_a^{-1} - n_c^{-1}|$ for out-of-plane birefringence and $\Delta n^{-1} = |n_b^{-1} - n_c^{-1}|$ for in-plane birefringence, of $NbOI_2$ along with those of other highly birefringent semiconductors and insulators. Plotting the inverse difference minimizes bias from absolute refractive-index magnitude. For non-layered three-dimensional (3D) materials, the out-of-plane birefringence is reported as the maximum $\Delta n^{-1}$ among all pairs of optical axes. In Fig. 1c, vdW materials[26–29,36] exhibit $\Delta n^{-1}$ an order of magnitude larger than archetypal anisotropic 3D crystals[37–40]. Even within the vdW family, $NbOI_2$ features record anisotropy both out-of-plane (Fig. 1c) and in-plane (Fig. 1d). Collectively, these properties enable very large optical SOI and nonlinear effects in $NbOI_2$.

We focus on natural slab waveguides of $NbOI_2$ on glass substrates prepared by mechanical exfoliation (Fig. 1e). The large dielectric contrast at both air/$NbOI_2$ and glass/$NbOI_2$ interfaces naturally supports low-loss guided modes inside $NbOI_2$ (inset in Fig. 1e). In Fig. 1f, a linearly polarized pump beam ($E$ = 1.3 eV) is focused onto the lower edge. The edge acts as a coupler that scatters far-field light into the $NbOI_2$ waveguide. The guided light propagates over more than 35 µm in the slab, outcoupling at the output (upper) edge. Bright features in Fig. 1e show the scattered pump light at the input and output edges, with negligible leakage between the two edges. Notably, we observe clear transverse splitting of the excitation light into two beams (illustrated in Fig. 1e), which emerge as two separate beams at the outcoupling edge. Using Stokes parameter imaging (Fig. 1g, see methods for details), we find that the outcoupled light exhibits a high degree of circular polarization (i.e. pseudospin purity), with $S_3$ = 0.95 (red) and -0.73 (blue) for the left and right beams, respectively. These results clearly indicate that beam splitting at the interface is subject to optical spin-momentum locking.[8,9] We return to a detailed analysis of beamsplitting and spin currents through spatiotemporal imaging below.

An outstanding challenge in integrated optics is weak nonlinearities, precluding active switching and wavelength conversion combined with beamsplitting and polarization control in the same optical element. Metal oxyhalides are uniquely suited to address this challenge due to their exceptionally large $\chi^{(2)}$.[10,12,34] Indeed, we observe strong second harmonic (SH) generation at 2.6 eV generated by the guided fundamental wave (FW). As shown in Fig. 1h, the SH light follows the same transverse trajectories as the FW in Fig. 1f. However, the SH light remains linearly polarized along the *b*-axis (linear polarization degree $|S_1| \sim 0.7$) because the largest $\chi^{(2)}$ in $NbOI_2$ is aligned along the polar *b*-axis (Fig. 1a). The Stokes vectors of both outcoupled FW and SH light are shown on the Poincaré sphere in Fig. 1i, with additional Stokes measurements and geometries reported in Supplementary Note 2. Thus, under linearly polarized excitation of the waveguide modes, $NbOI_2$ slabs act as a spin-splitter for FW light and as an efficient directional SH converter, enabling a unique degree of control over waveguided light.

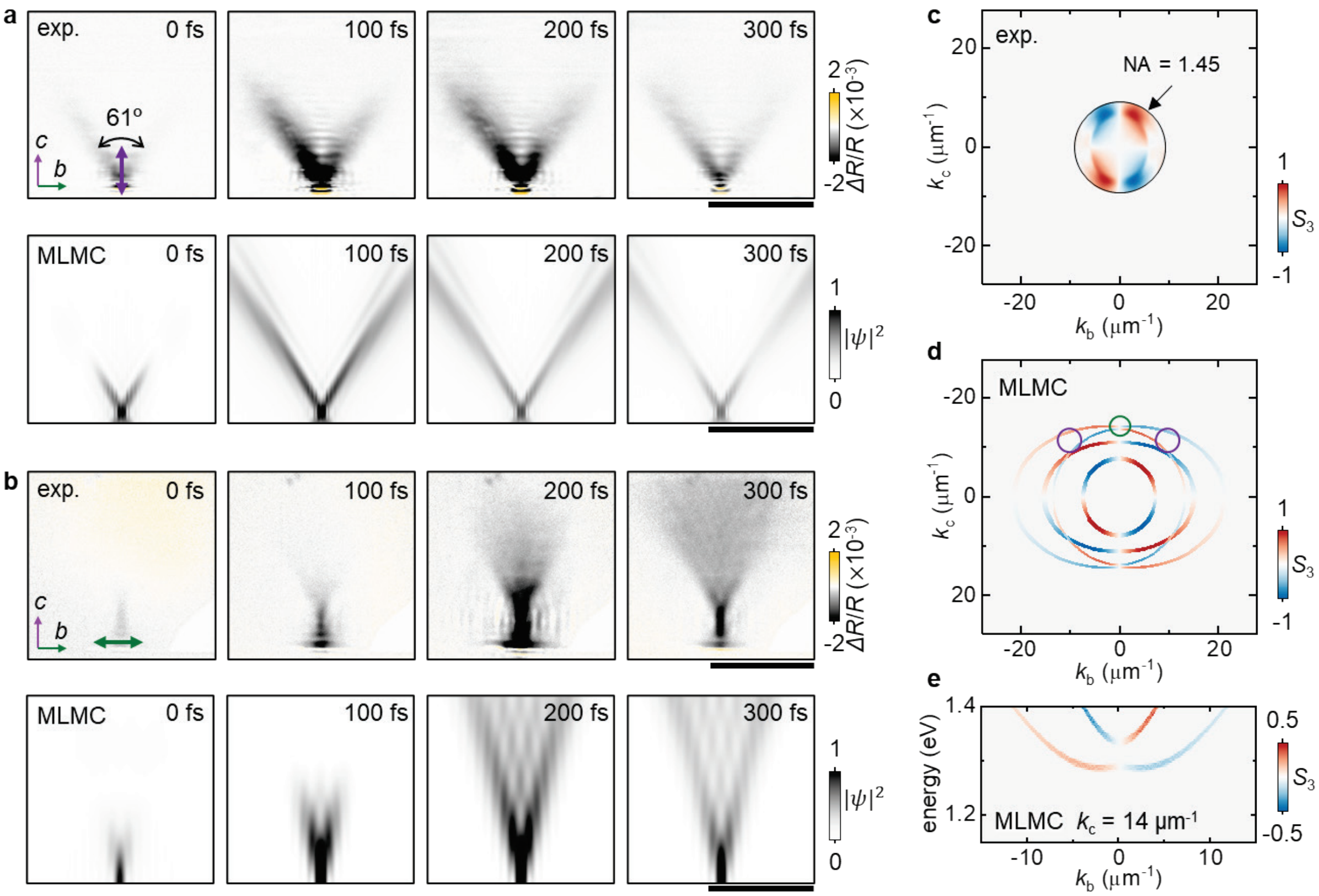

**Figure 2. Spin splitting and the waveguide spin Hall effect. a,b**, Spatiotemporal imaging of waveguided light propagation with $c$-polarized pump excitation (**a**, purple arrow) and $b$-polarized pump excitation (**b**, green arrow). The upper rows of each panel are experimental results; the lower rows show quantum dynamics simulations using the MLMC model. Scale bars are 10 μm. **c**, Experimental $S_3$-resolved isofrequency contour at $E = 1.3$ eV under $c$-polarized illumination. The experimentally accessible momentum range is limited by the far-field light cone (objective numerical aperture NA = 1.45, black line). **d**, Simulated $S_3$ isofrequency contour at $E = 1.3$ eV from the MLMC Hamiltonian. The corresponding in-plane momenta of waveguided modes in panels **a** and **b** are marked by purple and green circles, respectively. **e**, Simulated $S_3$ energy-momentum dispersion at $k_c = 14\ \mu m^{-1}$ from the MLMC Hamiltonian.

To understand the behavior of light inside $NbOI_2$ waveguides, we turn to a recently-developed pump-probe imaging approach leveraging the dynamic Stark effect to track light beyond the total internal reflection barrier in waveguides[30] (Methods and Extended Data Fig. 1). Briefly, a short pump pulse propagating in the waveguide exerts a transient and local change in the materials' polarizability. We track the spatiotemporal evolution of this change in femtosecond pump-probe microscopy, imaging the pump wavepacket as it propagates through the waveguide.[30] Fig. 2a displays the resulting images for the $NbOI_2$ slab under the same conditions as Fig. 1f ($b$- edge excitation with a pump polarized along the *c*-axis), providing direct access to the temporal evolution of the FW inside the waveguide. Fig. 2a clearly shows that the beam immediately splits in two at the waveguide interface, analogously to double refraction at birefringent interfaces. The two beams propagate at 41% of light speed (Extended Data Fig. 1) with nearly equal intensities and a large angular separation of 61°, remarkably allowing full resolution of individual beams within 1 μm of the edge.

**Optical spin currents through the waveguide optical spin Hall effect**

The propagation of light inside the waveguide is dramatically different depending on which crystal edge and which incident polarization is used (Extended Data Fig. 1d-g). Most notably, beam splitting at the edge occurs only when the excitation polarization is perpendicular to the principal axis (defined by the excited edge), leading to excitation of two waveguide modes and double refraction-like behavior as shown in Fig. 2a. Instead, when the polarization is along the principal axis (e.g. *b*-polarized excitation of the *b*-edge), we excite a single eigenmode of the waveguide. This scenario is shown in Fig. 2b (upper row). Here, the beam at first propagates perpendicular to the edge, i.e. with a wavevector $k_b = 0$. However, as the beam propagates, it exhibits an obvious broadening effect, with photon currents deflecting laterally on both sides. As shown below, this beam splitting during coherent light propagation is a characteristic signature of the OSHE, which deflects photons of opposite helicities in opposite directions.

To quantitatively model the observed behavior, we develop a fully microscopic light–matter coupling (MLMC) Hamiltonian[43] (Supplementary Note 3) that captures biaxial anisotropy, strong light-matter interactions, and the multimodal nature of the waveguide – features that are difficult to capture in prevalent phenomenological treatments of optical SOI. Comparison of our model with experimentally measured modal dispersions are in good agreement (Extended Data Fig. 2a-d). The isofrequency contours in the circular polarization basis at 1.3 eV are shown in Fig. 2d. The plot concurs with the cloverleaf pattern in experimental $S_3$ contours in Fig. 2c measured with angle-resolved Stokes parameter imaging (Methods). Note that the experimental measurements are limited to the passband of the far-field objective, with a numerical aperture of 1.45. Strong spin-momentum locking is evidenced by $S_3$ values approaching unity, and the cloverleaf pattern indicating that for finite values of $k_i$, $S_3(k_j) = -S_3(-k_j)$, where *i, j* denote different crystal axes, and *k* denotes in-plane momentum.

The contrasting transport behavior between Figs. 2a and 2b is best understood by inspecting the isofrequency contours in Fig. 2c-e. The strong dielectric anisotropy of $NbOI_2$ naturally induces pronounced optical SOI and lifts the degeneracy of cavity modes through large polarization splitting. In Fig. 2a, photoexcitation perpendicular to the edge is guided along two modes (purple circles in Fig. 2d) that undergo rapid spin precession due to this polarization splitting.[21,44] However, in Fig. 2b, photoexcitation along the principal axis populates a modal crossing, where the effective index for different polarization modes are degenerate (green circle in Fig. 2d, magnified at $k_c = 14\ \mu m^{-1}$ in Fig. 2e). These so-called diabolical points (DP) occur when the dielectric anisotropy compensates for waveguide-induced modal dispersion, suppressing spin precession and allowing purely geometric effects such as the OSHE to emerge. We model the resulting spatiotemporal evolution for these two different excitation conditions through quantum dynamics simulations (Supplementary Note 3). The results are shown in the lower rows of Figs 2a and 2b, displaying very good agreement with experiments. A high-resolution calculation of the energy-momentum dispersion in Fig. 2e shows a small avoided crossing arising from strong light-matter coupling (polariton formation) in $NbOI_2$, which leads to matter-mediated interactions even between modes of opposite parity. These interactions are likely overestimated in the model due to the sensitivity to the exact choice of the dielectric function (Supplementary Note 3.3).

Spin-resolved simulations in Fig. 3a-c reveal the interplay of OSHE and spatial propagation in highly anisotropic biaxial waveguides. We choose an energy for the MLMC model with weaker light-matter interactions ($E = 1$ eV) to cleanly illustrate light polarization dynamics. Fig. 3a shows an isofrequency contour near the DP. Figure 3b shows guided modes launched with finite transverse momentum from edge refraction (purple circles in Fig. 3a, corresponding to conditions in Fig. 2a). The simulations show rapid spin oscillations due to anisotropy-induced splitting of the polarization eigenmodes. We note that the different Poynting vectors of oppositely polarized modes eventually leads to spatial separation of overall spin-polarized currents. In contrast, at the DP (green circle in Fig. 3a, corresponding to conditions in Fig. 2b), opposite photon helicities experience opposite transverse drift, leading to clean spatial separation of the two light helicities (Fig. 3c). These results show that strong optical SOI in $NbOI_2$ drives robust spatial separation of photon helicities both at the DP and away from it.

To frame our findings in terms of spin–orbit coupling parameters widely used by the field, we project the dominant dispersion branches onto an effective Rashba–Dresselhaus formalism:[8,9,19,44,45]

$$H_{\mathrm{RD}} = H_{\mathrm{c}}\mathbb{I}_{2\times2} + \vec{G}\cdot\vec{\sigma}\,, \qquad (1)$$

where $H_c$ is the photon Hamiltonian of degenerate cavity modes, $\mathbb{I}_{2\times2}$ is the $2 \times 2$ identity matrix, and $\vec{\sigma}$ is the Pauli matrix representation of the polarization states of photons. The effective magnetic field $\vec{G} = [-\alpha+\beta k^2\cos(2\varphi), \beta k^2\sin(2\varphi), \delta_z]$ depends on in-plane anisotropy $\alpha$, out-of-plane TE-TM splitting as $\beta k^2$, and the in-plane wavevector angle $\varphi$. The effective Zeeman splitting $\delta_z$ is zero in the absence of an external field.[46–48] For the lowest-order waveguide mode in our $NbOI_2$ slab, our fits yield parameters of $\alpha = 0.204$ eV, and $\beta = 1.02$ meV·µm$^2$. Fig. 3d shows the in-plane projection of effective magnetic field $\vec{G}$. At the DP (at $|k_c| \approx 14$ µm$^{-1}$), $\vec{G}$ vanishes, and a topological singularity emerges (yellow spots in Fig. 3d). At this point, the Berry curvature diverges, allowing for spatial separation of pure spin currents through geometric phase without spin precession.

The interplay of spin oscillations and transverse Hall drift produces mode-dependent transverse beam broadening during propagation, in addition to beam splitting. Transverse Hall drift leads to large beam broadening at the DP in Fig. 2b. Conversely, away from the DP, rapid spin precession leads to minimal divergence and well-defined, narrow beams in Fig. 2a. These two regimes are advantageous for different purposes: long-lived pure spin currents are best achieved at the DP, whereas low-loss interconnects and beamsplitters benefit from low divergence. Fig. 3e quantitatively confirms this interplay between OSHE and spin precession on beam divergence.

Beam width from dynamical simulations (grey) using our MLMC model show excellent agreement with experiments (yellow). This diversity of behavior observed within the same waveguide enables truly multifunctional polarization manipulation in a single nanoscale optical element.

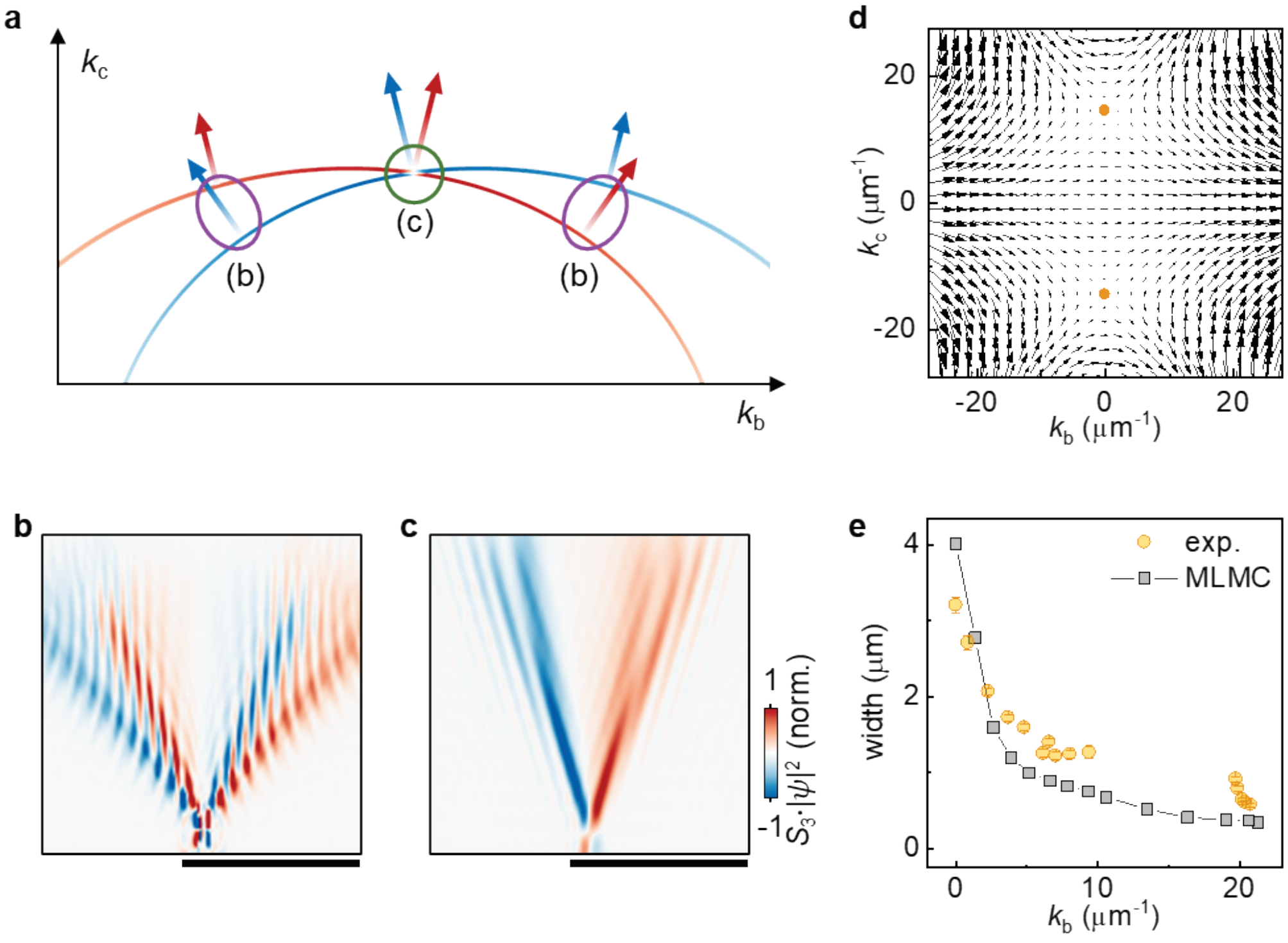

**Figure 3. Interplay of spin precession and spin-Hall drift. a**, Magnified isofrequency contour near a diabolical point at $E$ = 1.0 eV. **b**, $S_3$ dynamical propagation calculated using the MLMC Hamiltonian for splitting guided modes with rapid spin-precession. **c**, Same as (b) at DP. **d.** In-plane projection of effective magnetic field $\vec{G}$. The DP is indicated by yellow circles. **e**. Beam width of the guided modes as a function of $k_b$ from DP. Simulated beam widths (grey) and experimental beam widths (yellow) are measured 7 μm from excitation. Error bars denote the standard deviation from the fits.

## On-demand beam steering in $NbOI_2$ waveguides

We now demonstrate programmable beam manipulation through polarization-controlled optical beam splitting and steering in $NbOI_2$ slab waveguides (Fig. 4a). We track propagating beams inside the waveguide using ultrafast imaging as a function of the pump polarization orientation $\psi$ (with respect to the excited $b$-edge). Fig. 4b summarizes the polarization-dependent splitting behavior, extracted from profiles of the beam intensities 5 μm along the $c$-axis from the excitation edge (dashed rectangles in insets). Equal-intensity beam splitting occurs for $\psi = \pi/2$ ($c$-axis polarization). Rotating the pump polarization away from the crystal axes induces asymmetric splitting and propagation: one beam increases in intensity and steers towards the edge normal, while the other beam exhibits the opposite trend. The gray insets of Fig. 4b at $\psi$ = π/4 and 3π/4 illustrate this asymmetric behavior. The polarization dependence of the intensity splitting between left and right beams is displayed in Fig. 4c, emphasizing the asymmetry. As aforementioned, at $\psi = 0, \pi$ ($b$-axis polarization), the pump couples into the DP, leading to edge-normal propagation and transverse drift into spin-split currents (green inset). The beam steering capability of the waveguide is also captured by plotting the gain or loss of transverse momentum $k_b$ in Fig. 4d. Note that total momentum $k_b$ must be conserved across the $b$-axis refraction interface, as verified by plotting the total transverse momentum (Fig. 4e).

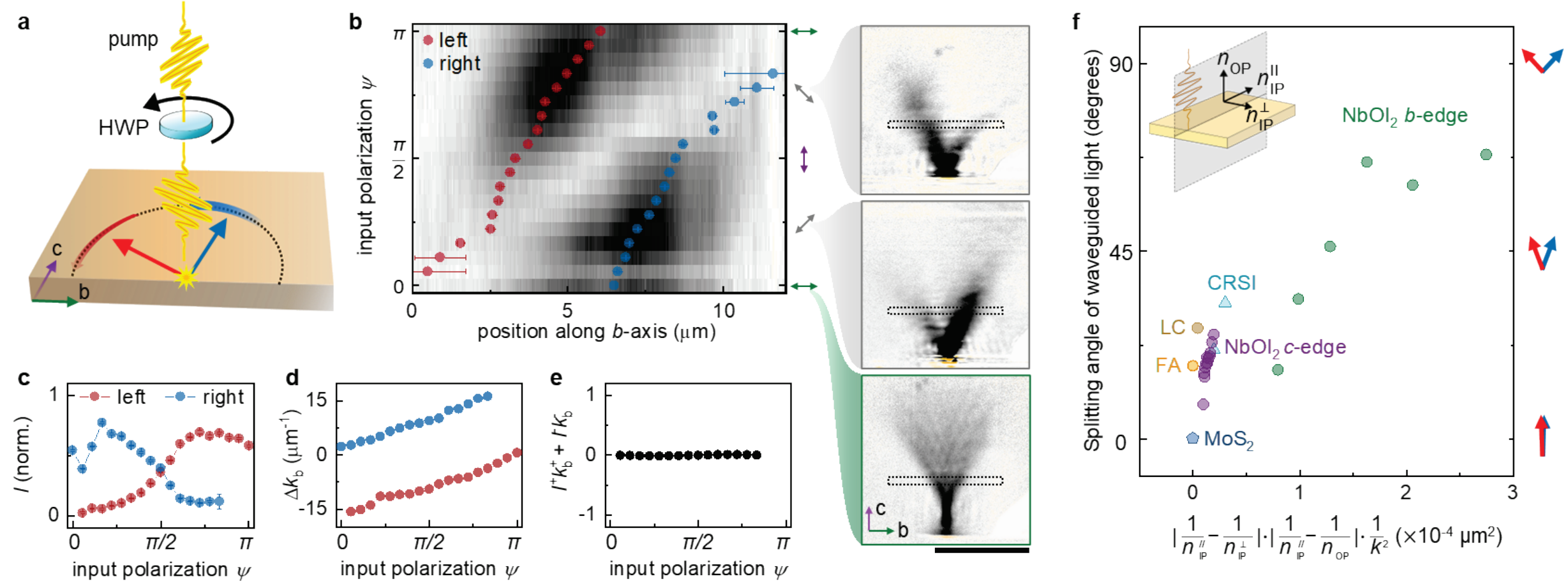

**Figure 4. $NbOI_2$ waveguides as beam steerers. a**. Schematic of polarization-dependent beam steering in $NbOI_2$ waveguides. The pump polarization orientation is controlled by a half-waveplate (HWP). **b**. Intensity profiles of waveguided light measured 5 μm away from the excitation edge at 300 fs pump-probe time delay (dashed rectangle in the insets). The centroids of the left (red circles) and right (blue circles) modes are obtained from Gaussian fits to the intensity profiles. Error bars are the fit standard deviations. The insets show the propagation for pump polarizations $\psi = 3\pi/4$, $\pi/4$, and 0 relative to the edge. **c–e**. Polarization-dependent evolution of beam splitting and steering. The intensities of the split beams in (c) are extracted from (b). Transverse momenta $k_b$ in (d) are determined from the measured propagation directions in panel (b) and group velocities (Extended Data Fig. 1). The total transverse momenta in (e) are calculated by summing over the products of transverse momenta $k_b$ and intensities $I$ for each beam. **f**. Beam-splitting angles across different optical SOI systems. The horizontal axis is the empirical anisotropy coefficient. The splitting angles for $NbOI_2$ excited at the $c$-edge (purple circles), $NbOI_2$ at the $b$-edge (green circles), $CsRe_6Se_8I_3$ (CRSI, cyan triangle), and $MoS_2$ (blue pentagon) are measured in this work. Multiple circles for the same system indicate measurements at different energies. Splitting angles from the literature in OSHE systems $FAPbBr_3$ cavities (FA, yellow circle)[21] and liquid crystals hybrids cavities (LC, brown circle)[22] are also plotted.

Finally, we analyze beam steering capabilities under different excitation conditions in $NbOI_2$ and across different material systems. Fig. 4f plots the measured beam-splitting angles in the $NbOI_2$ waveguide for $b$- and $c$-edge excitation at energies ranging from 1.2 to 1.7 eV. We observe a linear dependence of the splitting angle on the empirical total anisotropy coefficient $\alpha \times$ $\beta \propto |\frac{1}{n_{IP}^{\parallel}} - \frac{1}{n_{IP}^{\perp}}||\frac{1}{n_{IP}^{\parallel}} - \frac{1}{n_{OP}^{\perp}}| \cdot \frac{1}{k^2}$, where $n_{IP}^{\parallel}$, $n_{IP}^{\perp}$, and $n_{OP}$ are the refractive indices along the three principal axes relative to the incidence plane (Fig. 4f inset). This coefficient reflects the combined roles of in-plane and out-of-plane birefringence, as introduced in Figs. 1c-d. The waveguided light launched along the $b$-edge of $NbOI_2$ (green circles in Fig. 4f) exhibits a larger splitting angle than that launched along the $c$-edge (purple circles in Fig. 4f) owing to the much stronger dielectric contrast with the $a$-axis.

We extend the correlation of splitting angle with anisotropy coefficient to other vdW waveguides and planar cavities in Fig. 4f. We measured the spin splitting angles in natural waveguides made from $CsRe_6Se_8I_3$ (CRSI)[49,50], an anisotropic vdW superatomic semiconductor, and $MoS_2$, a transition metal dichalcogenide semiconductor (Extended Data Fig. 4). We further include results from recent studies on OSHE in $FAPbBr_3$ cavities (FA, reported by Shi *et al.*[21]), and liquid crystal hybrid cavities (LC, reported by Liang *et al*[22]). We confirm that the trend linking splitting angle to anisotropy coefficient is maintained across all material platforms in Fig. 4f. Among reported systems, $NbOI_2$ exhibits the largest splitting angle of up to 70° within sub-micrometer propagation due to the strongest anisotropy, enabling sub-micrometer beam splitting and a very strong response to polarization control. Comparable optical SOI has only been achieved in engineered metasurfaces, such as hyperbolic metamaterials[51,52] and helical topological polaritons[53,54]. Our results thus demonstrate that optimizing for extreme material anisotropy in simple photonic structures can deliver optical SOI performance comparable or exceeding that of complex metasurface architectures. Overall, $NbOI_2$ waveguides act as remarkably efficient, broadband, and controllable on-chip frequency converters, beamsplitters, steerers, and polarizers.

**Conclusion**

In summary, we demonstrate strong and tunable optical SOI in highly anisotropic and chip-compatible waveguides. Unlike previous approaches that rely on engineered photonic structures, the record dielectric anisotropy of the ferroelectric semiconductor $NbOI_2$ provides an ideal playground for the exploration of optical SOI. We show that the large refractive index and strong nonlinearities inherent to this material enables natural low-loss waveguides, strong light-matter interactions, and large optical nonlinearities including harmonic conversion. We successfully demonstrate guided light transport over tens of micrometers, and very large beamsplitting angles allowing spatial resolution of spin-split photon currents within sub-micrometer propagation distances. Beam splitting and steering is controlled over a broad frequency range with exquisite sensitivity through input light polarization. This capability should be of key interest in remote sensing technology (e.g. light detection and ranging - LiDAR), and should enable ultrafast control of waveguide propagation for dense photonic networks such as optical computers. We further show that optical spin precession can be suppressed when anisotropy balances mode dispersion. In this regime, we observe transverse drift of waveguided light and pure spin current separation through the OSHE. Our materials-driven approach enables optical functionality to emerge from intrinsic crystal properties and identifies highly anisotropic vdW waveguides as compelling platforms for nonlinear spin-optronic devices operating at room temperature in dense photonic environments.

**Data availability**

The data that support the findings of this article are in the main text and supplementary information. Raw data files are available from the authors. The transfer matrix code is openly available on Zenodo at https://doi.org/10.5281/zenodo.18532292.

## Methods

### $NbOI_2$ waveguide preparation

Bulk $NbOI_2$ single crystals were synthesized via chemical vapor transport as described in Choe *et al.*[11]. Thin flakes were mechanically exfoliated from the bulk crystal and transferred onto glass substrates (Thorlabs, CG15CH2). To prepare large-area $NbOI_2$ slab waveguides cleaved along well-defined crystallographic edges, low-adhesion tape (Nitto, SPV-224PR-MJ) was used for a single-step exfoliation, and the tape was peeled off along the crystallographic edge direction.

### Ultrafast imaging

A diode-pumped ultrafast Ytterbium-doped photonic crystal fiber amplifier system (Amplitude Tangerine-SP, 50 W, 1030 nm fundamental, 10 MHz repetition rate) is used as the laser source. Two independent supercontinuum white-light beams for pump and probe are generated separately by focusing the 1030 nm fundamental onto YAG windows (EKSMA, 555-712, ø 12.7 mm, 5 mm). The white-light beams are compressed using two prism-pair compressors (EKSMA, LAK21, 25.4 mm × 50.8 mm) to compensate for the group-delay dispersion of the imaging system. The pump and probe beams are directed to a high numerical-aperture objective (Olympus Plan Apo 100x, 1.4 NA oil immersion). The pump beam is spatially filtered using a pinhole assembly to produce a Gaussian profile enabling diffraction-limited excitation of the sample. Widefield probe illumination is achieved by focusing the probe beam onto the back-focal plane of the objective. The back-scattered light from the sample together with the reflected probe light from the sample-substrate interface forms interferometric scattering images[55–57] on a CMOS camera (Blackfly S USB3, BFS-U3-28S5M-C). Experimental setup details are provided in Supplementary Note 1.

### Full Stokes parameter imaging

Full Stokes parameter imaging is performed using a refractive microscope integrated with a polarization-analyzing component placed at either the real-space image plane or the back-focal plane, enabling spatial- or momentum-resolved polarization mapping. The back illumination source is a stabilized tungsten-halogen lamp (Thorlabs, SLS201L, 360-2600 nm). The polarization analyzer consists of a rotating quarter-wave plate (QWP) followed by a fixed linear polarizer, following the procedure in Schaefer *et al.*[58] Experimental setup details are provided in Supplementary Note 1. The QWP introduces a relative phase shift between orthogonal polarization components of the reflected field. By rotating the QWP, the detected intensity at each pixel is:

$$I(\vec{r}, \theta) = \frac{1}{2}[\, S_0(\vec{r}) + S_1(\vec{r})\cos^2 2\theta + S_2(\vec{r})\cos 2\theta \sin 2\theta + S_3(\vec{r})\sin 2\theta \,] \tag{2}$$

where $\vec{r}$ is the detector coordinates (real space or momentum space), $\theta$ is the angle of the QWP fast axis, and $S_n(\vec{r})$ are the spatially resolved Stokes parameters. For each measurement, we acquire 18 images with the QWP fast axis rotating from 0° to 170° in 10° steps. The Stokes parameters at each pixel are extracted from the angular Fourier components of $I(\vec{r}, \theta)$. We reconstruct the Stokes parameter images using the following linear conversion matrix:

$$\begin{pmatrix} S_0(\vec{r}) \\ S_1(\vec{r}) \\ S_2(\vec{r}) \\ S_3(\vec{r}) \end{pmatrix} = \frac{4}{m} \begin{pmatrix} 1 & 0 & -1 & 0 \\ 0 & 0 & 2 & 0 \\ 0 & 0 & 0 & 2 \\ 0 & 1 & 0 & 0 \end{pmatrix} \begin{pmatrix} 1/2 \sum I_n \\ \sum I_n \sin 2\theta_n \\ \sum I_n \cos 4\theta_n \\ \sum I_n \sin 4\theta_n \end{pmatrix} \quad (3)$$

where $m$ is the number of images for a full rotation measurement. The Stokes parameters are normalized to $S_0(\vec{r})$ for imaging reconstruction. For waveguided light collected from the outcoupling edge, the measured Stokes vectors are modified by edge refraction. We apply a Mueller matrix correction describing the edge refraction to recover the intrinsic Stokes parameters of the waveguided light (Supplementary Note 2).

**Transfer matrix method**

The polarization-resolved mode dispersions of $NbOI_2$ waveguides beyond the light line in Extended Data Fig. 2 are calculated using a transfer-matrix method. Multilayer stacks were specified by the material identity and thickness of each layer. At oblique incidence, the dielectric tensor of each material was rotated into the local reflection basis as $\varepsilon_{\mathrm{loc}}(\omega, \vec{k}_{\parallel}) = R^{\mathrm{T}}\varepsilon(\omega)R$, where $\omega$ is the optic frequency, $\varepsilon(\omega)$ is the dielectric function measured along crystallographic direction, and $\vec{k}_{\parallel}$ is the in-plane wavevector. The local reflection polarization basis $(\hat{s}, \hat{p}, \hat{z})$ is defined as $\hat{s} \perp \vec{k}_{\parallel}$, $\hat{p} \parallel \vec{k}_{\parallel}$, $\hat{z}$ is the out-of-plane direction, and $R$ transforms the laboratory frame to this local basis. The incident field was specified by a Jones vector with defined polarized field $E_{\mathrm{s}}$, $E_{\mathrm{p}}$, and a relative phase. At each interface, a 2×2 Jones reflection matrix was constructed from the Fresnel coefficients calculated using the local dielectric tensor. The total reflection matrix of the multilayer structure was obtained by recursively cascading interface reflection matrices with propagation phases through the stack and subsequently rotating the reflected field back into the laboratory frame. The energy-momentum dispersion along a specified momentum direction is calculated from reflectance on an energy-momentum mesh. The energy cross-section dispersion is calculated on a uniform in-plane momentum mesh. The polarization state of the reflected light was characterized using the Stokes parameters ($S_1$, $S_2$, $S_3$) normalized to total reflectance $S_0$.

**Microscopic light-matter coupling Hamiltonian for multimode anisotropic waveguides**

To numerically simulate optical SOI and elucidate the observed transport properties, we employ a microscopic light-matter model that captures the anisotropic light-matter coupling in a multimode cavity. In this model, we consider the dielectric function in a Lorentz-oscillator model with $N_{\mathrm{d}}$ transition dipoles:

$$\varepsilon(\omega) = \varepsilon_{\infty} + \sum_{\ell}^{N_d} \frac{\epsilon_{\ell} \left| \vec{\mu}_{\ell} \right|^2}{\epsilon_{\ell}^2 - \omega^2}$$

where $\omega$ is the optical frequency, $\epsilon_{\ell}$ are excitonic transition energies, and $\vec{\mu}_{\ell}$ are the corresponding exciton transition dipole moments. We fit this model to experimentally measured dielectric functions and extract the set of $\epsilon_\ell$ and $|\vec{\mu}_\ell|^2$ along $b$- and $c$-axes. For a planar cavity with photonic confinement along $z$ (out-of-plane direction), the three-dimensional photonic modes wavevector is $\vec{k}(q) = \vec{k}_\parallel + k_{q,z}\vec{z}$, where $k_{q,z} = \frac{q\pi}{L}$ ($q$ is a positive integer) and $\vec{k}_\parallel$ is the in-plane momentum. The corresponding photonic mode frequency for both TE and TM modes is $\omega_{q,k} = \frac{c}{\eta}|\vec{k}(q)| = \frac{c}{\eta}\sqrt{|\vec{k}_\parallel|^2 + k_{q,z}^2}$, where $c$ is the speed of light in free space and $\eta$ is dielectric background refractive index. The full MLMC Hamiltonian can be block-diagonalized with respect to $\vec{k}(q)$, and each block is written as:[59,60]

$$\hat{H}_{LM}(q,\boldsymbol{k}) = \begin{pmatrix} H_{ex}(\boldsymbol{k}) & H_I(q,\boldsymbol{k}) \\ H_I^*(q,\boldsymbol{k}) & H_{ph}(q,\boldsymbol{k}) \end{pmatrix}$$

where $H_{ex}(\boldsymbol{k})$ = diag$\{\epsilon_{1,k}, \ldots \epsilon_{N,k}\}$ is the excitonic block, $H_{ph}(q,\vec{k})$ = diag$\{\omega_{q,k}, \omega_{q,k}\}$ is the photonic block containing the TE and TM modes, and $H_I^*(q,\boldsymbol{k})$ is an $N$×2$q$ matrix describing the corresponding exciton–photon couplings. For the quantum dynamic simulations, we solve the Schrodinger equation for the time evolution of the initial exciton-photon states at each $\vec{k}(q)$, and then evaluate the Stokes parameters using Pauli matrices. Further details of the MLMC, quantum dynamics and effective Rashba-Dresselhaus Hamiltonian are discussed in Supplementary Note 3.

**Method References**

**Acknowledgements**
Work on waveguide propagation in $NbOI_2$ was supported by the Office of Naval Research under award N00014-25-1-2079; and by Programmable Quantum Materials, an Energy Frontier Research Center funded by the US Department of Energy (DOE), Office of Science, Basic Energy Sciences, under award DE-SC0019443. Instrument development for ultrafast waveguide imaging was supported by the Air Force Office of Scientific Research (AFOSR), under DURIP award FA9550-22-1-0500. Superatomic material synthesis was supported by the AFOSR, under MURI grant number FA9550-25-1-0288. Stokes parameter imaging was supported by the National Science Foundation under award CHE-2203844. M.D. also acknowledges support from the Alfred P. Sloan Foundation through a Sloan Fellowship, and from the Camille and Henry Dreyfus Foundation through a Camille Dreyfus Teacher-Scholar award. A.M., A. A. and S.R.K. acknowledge support from the Texas A&M startup funds. S.R.K. and A.A. acknowledge computational support from the Advanced Cyberinfrastructure Coordination Ecosystem: Services & Support (ACCESS) program (allocations: PHY240260), under National Science Foundation grants #2138259, #2138286, #2138307,#2137603, and #2138296. V.Q-C. acknowledges support from the Columbia Quantum Initiative Postdoctoral Fellowship.

**Author Contributions**
D.X. and M.D. conceived and designed the experiment. D.X., V.G., Y.H., and A.K.L performed ultrafast microscopy experiments. D.X., J.A.T., and V.Q-C. performed the steady-state measurements. S.R.K., A.A., and A.M. developed the theory and performed the quantum dynamical simulations. C-Y. H., T.H., A.K.L, D.G.C, and E.L. synthesized the crystals. P.J.S., X.Y., X.R., A.M. and M.D. supervised the study. D.X., S.R.K., A.M. and M.D. wrote the article with input from all authors.

**Competing Interests**
The authors declare no competing interests

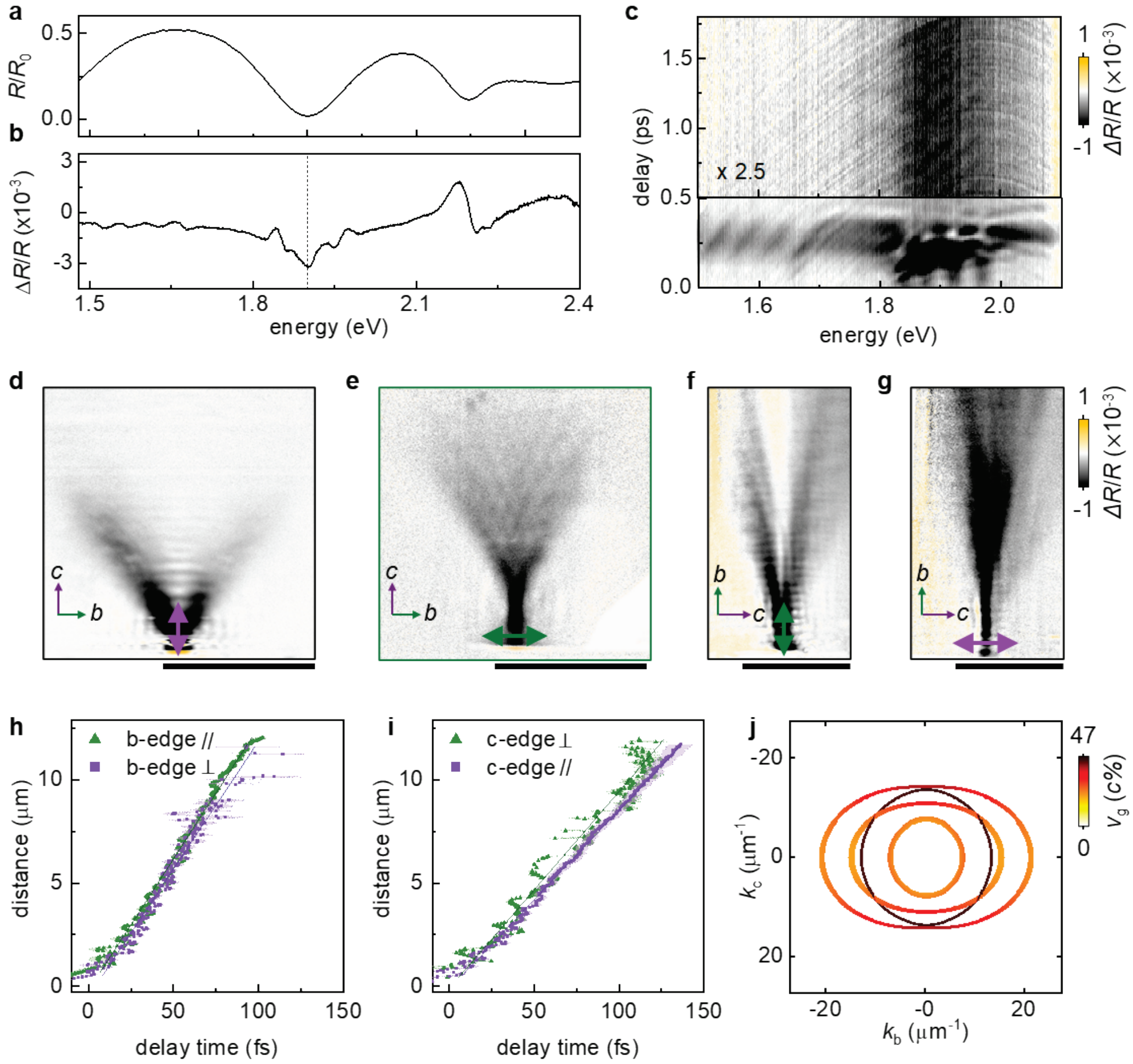

**Extended Data Fig. 1 | Ultrafast imaging of waveguided light propagation. a**, Reflectance of $NbOI_2$ slab waveguide at $k = 0$ under $c$-polarized illumination, normalized to reflectance from a silver film on glass ($R_0$). **b**, Transient reflectance spectrum $\Delta R/R$, following excitation of the edge at $E_{pump}$ = 1.3 eV, and measured with a probe located 1 μm away from the edge at 300 fs pump–probe delay with a $c$-polarized supercontinuum probe. The dashed line at $E_{probe}$ = 1.9 eV indicates the probe energy used for ultrafast imaging. **c**, Time-resolved $\Delta R/R$. The strong signal before 500 fs is dominated by the Stark effect near the cavity resonance, whereas the oscillatory feature appearing after 500 fs is assigned to a ferron oscillation at approximately 3 THz.[11] **d-g**, Snapshots of waveguided light propagation recorded at a pump–probe delay of 300 fs. Arrows indicate the pump polarization orientation. Scale bars are 10 μm. **h-i**, Waveguided light propagation traces extracted from the four pump-polarization configurations shown above. The measured group velocities are 41% and 43% of light speed for edge-perpendicular (purple in h) and edge-parallel (green in h) pumping at the $b$-edge, respectively, and 32% and 29% of light speed for edge-perpendicular (green in i) and edge-parallel (purple in **i**) pumping at the $c$-edge, respectively. Data are shown as mean ± one standard deviation from the fitting error. **j**, Group velocity of the waveguide modes at $E$ = 1.3 eV, calculated from the MLMC Hamiltonian.

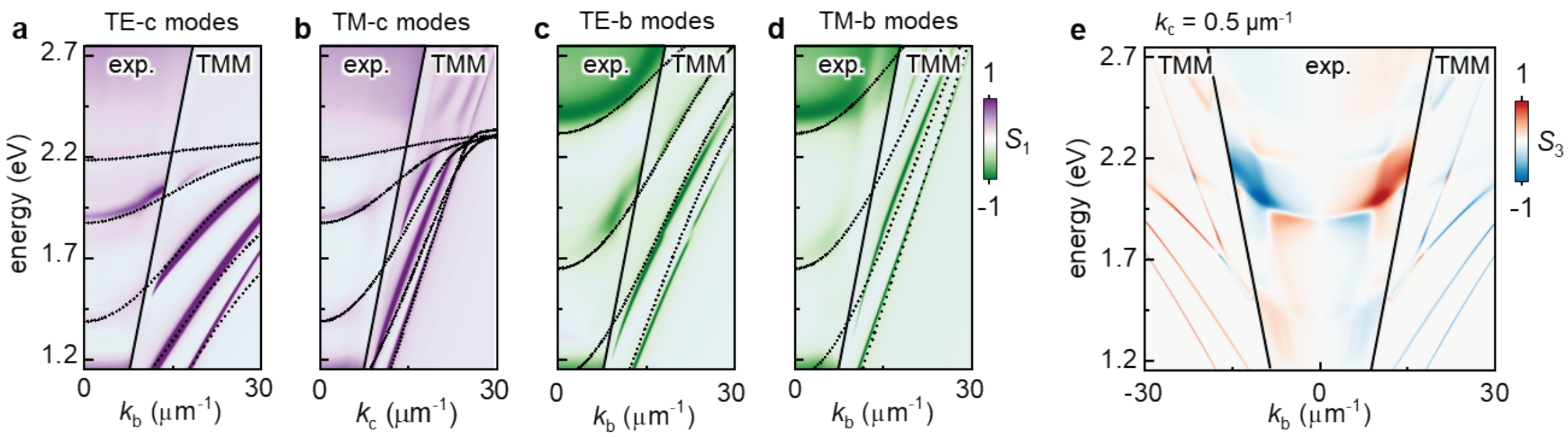

**Extended Data Fig. 2 | Slab waveguide dispersion with optical SOI. a–d**, $S_1$ Stokes parameter of $NbOI_2$ slab waveguide dispersions measured by angle-resolved reflectance spectroscopy. The experimentally accessible momentum range is limited by the numerical aperture (NA = 1.45, black lines). High-momentum dispersions beyond this range are obtained from the transfer matrix method (TMM). Modes polarized along the *c* and *b* axes exhibit $S_1 > 0$ (purple) and $S_1 < 0$ (green), respectively. Dashed curves indicate dispersions calculated from the MLMC Hamiltonian. **e**, Experimental $S_3$ at $k_c = 0.5$ μm$^{-1}$, measured under linearly polarized illumination along the *c*-axis. The dispersion beyond the NA is calculated using the TMM.

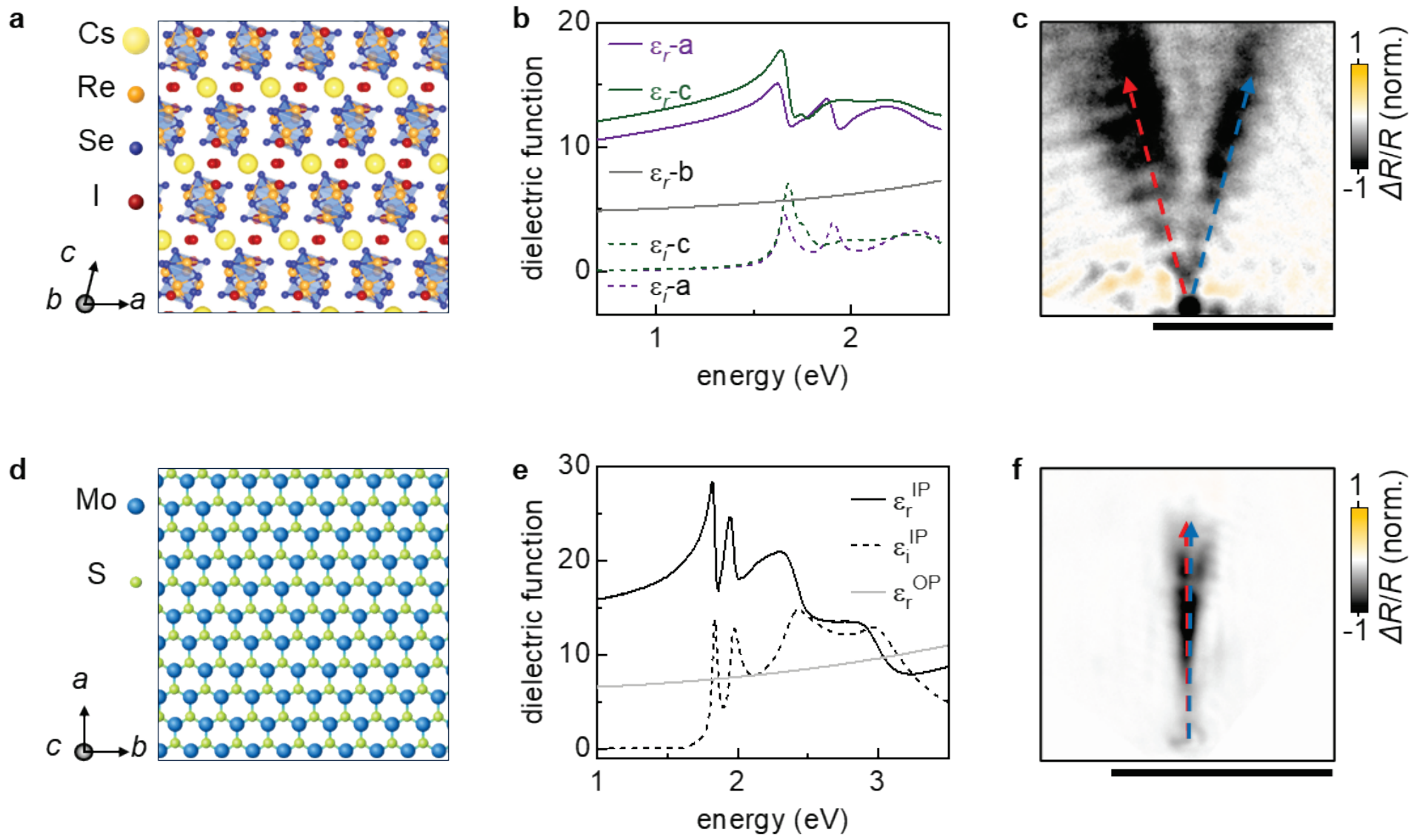

**Extended Data Fig. 3 | Beam splitting in anisotropic and isotropic van der Waals slab waveguides**. **a**, In-plane crystal structures of $CsRe_6Se_8I_3$. **b**, Dielectric function of $CsRe_6Se_8I_3$ **c**, Snapshot of waveguide propagation in $CsRe_6Se_8I_3$, with excitation at $E_{pump}$ = 1.3 eV and probing at the bandgap resonance at $E_{probe}$ = 1.8 eV edge at 300 fs pump–probe delay. **d–f**, Same as **a–c**, for $MoS_2$. Scale bars are 10 μm.

**Supplementary Information for:**

**Giant optical spin-orbit interactions in ferroelectric van der Waals waveguides**

Ding Xu[1†], Saeed Rahmanian Koshkaki[2†], Vicente Galicia[1], Chun-Ying Huang[1], Victoria Quiros-Cordero[1,3], Jakhangirkhodja A. Tulyagankhodjaev[1], André Koch Liston[1], Daniel G. Chica[1], Emma Lian[1], Amirhosein Amini[2], Yongseok Hong[1], Taketo Handa[1], P. James Schuck[3], Xiaoyang Zhu[1], Xavier Roy[1], Arkajit Mandal[2]*, Milan Delor[1]*

1. Department of Chemistry, Columbia University, New York, NY 10027, USA
2. Department of Chemistry, Texas A&M University, College Station, Texas 77843, USA
3. Department of Mechanical Engineering, Columbia University, New York, NY, 10027, USA

[†] Equal contribution

*mandal@tamu.edu; milan.delor@columbia.edu

## 1. Experimental setups for ultrafast imaging

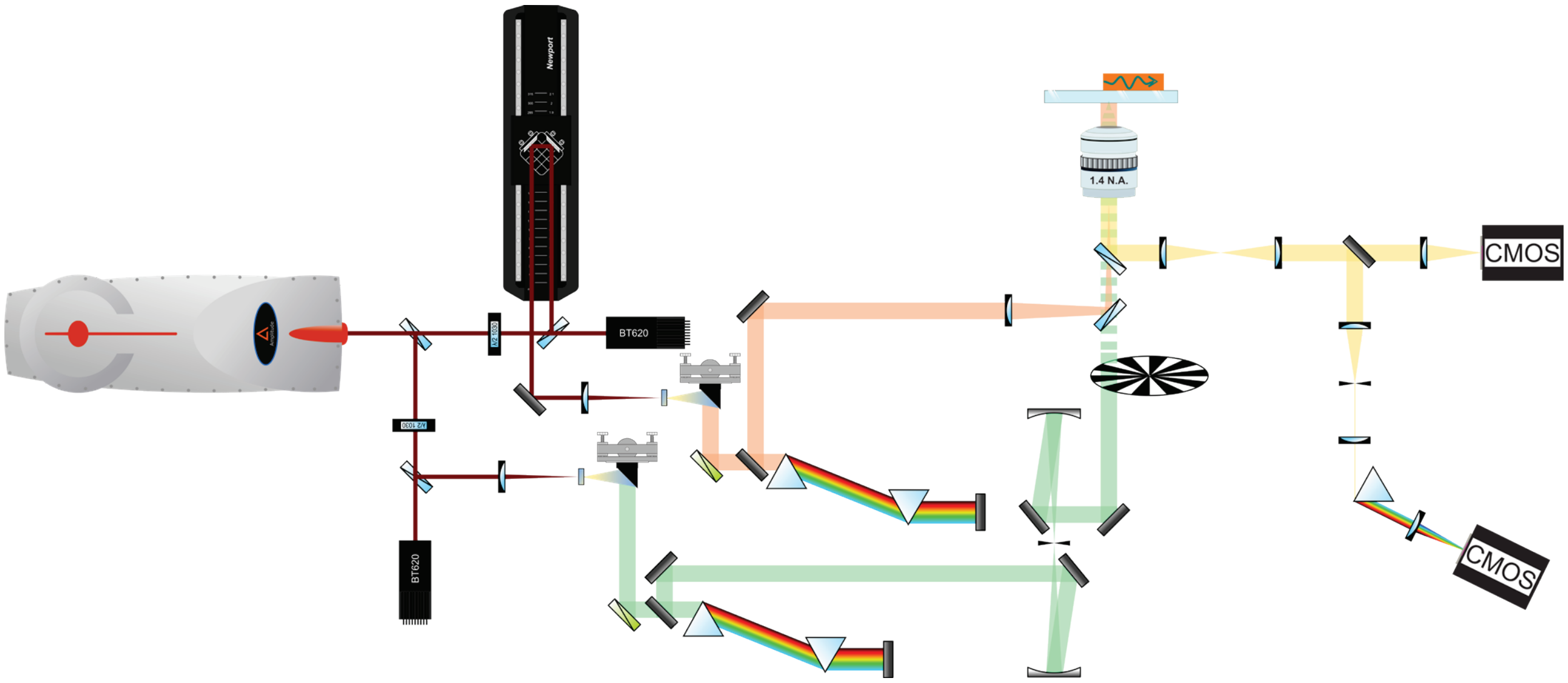

**Figure S1. Setup schematic for ultrafast waveguide imaging**. A diode-pumped ultrafast Ytterbium-doped photonic crystal fiber amplifier system (Amplitude Tangerine-SP, 50 W, 1030 nm fundamental, 10 MHz repetition rate) is used as the laser source. The output is split into two 3 W beams to generate independent supercontinuum white-light pulses for pump and probe. The probe-generation beam is routed through a mechanical delay line to control the pump-probe delay. Supercontinuum pulses spanning approximately 450-1000 nm were generated by focusing each fundamental beam onto a YAG window (EKSMA, 555-712, ø 12.7 mm, 5 mm) using a $f$ = 100 mm lens (EKSMA, Femtoline Thin AR Lenses). The resulting white-light pulses were spectrally filtered to ~ 50 nm narrow band pulses and pre-chirped using a prism-pair compressor (EKSMA, LAK21, 25.4 mm × 50.8 mm) to compensate for group-delay dispersion introduced by the imaging system. Pump and probe beams are directed to a high numerical-aperture objective (Olympus Plan Apo 100x, 1.4 NA, oil immersion). The pump beam is spatially filtered using a pinhole assembly to produce a Gaussian profile and enable diffraction-limited excitation. A mechanical chopper modulates the pump beam to generate alternating pump-on and pump-off pulse trains synchronized to the detector. Wide-field probe illumination is achieved by focusing the probe beam onto the back-focal plane of the objective lens using an $f = 300$ mm lens. Back-scattered light from the sample, together with the reflected probe field, forms interferometric scattering images on a CMOS camera (Blackfly S USB3, BFS-U3-28S5M-C). A flip mirror in the image path allows redirecting the reflected signal to a prism spectrometer for energy-dispersion measurements. s

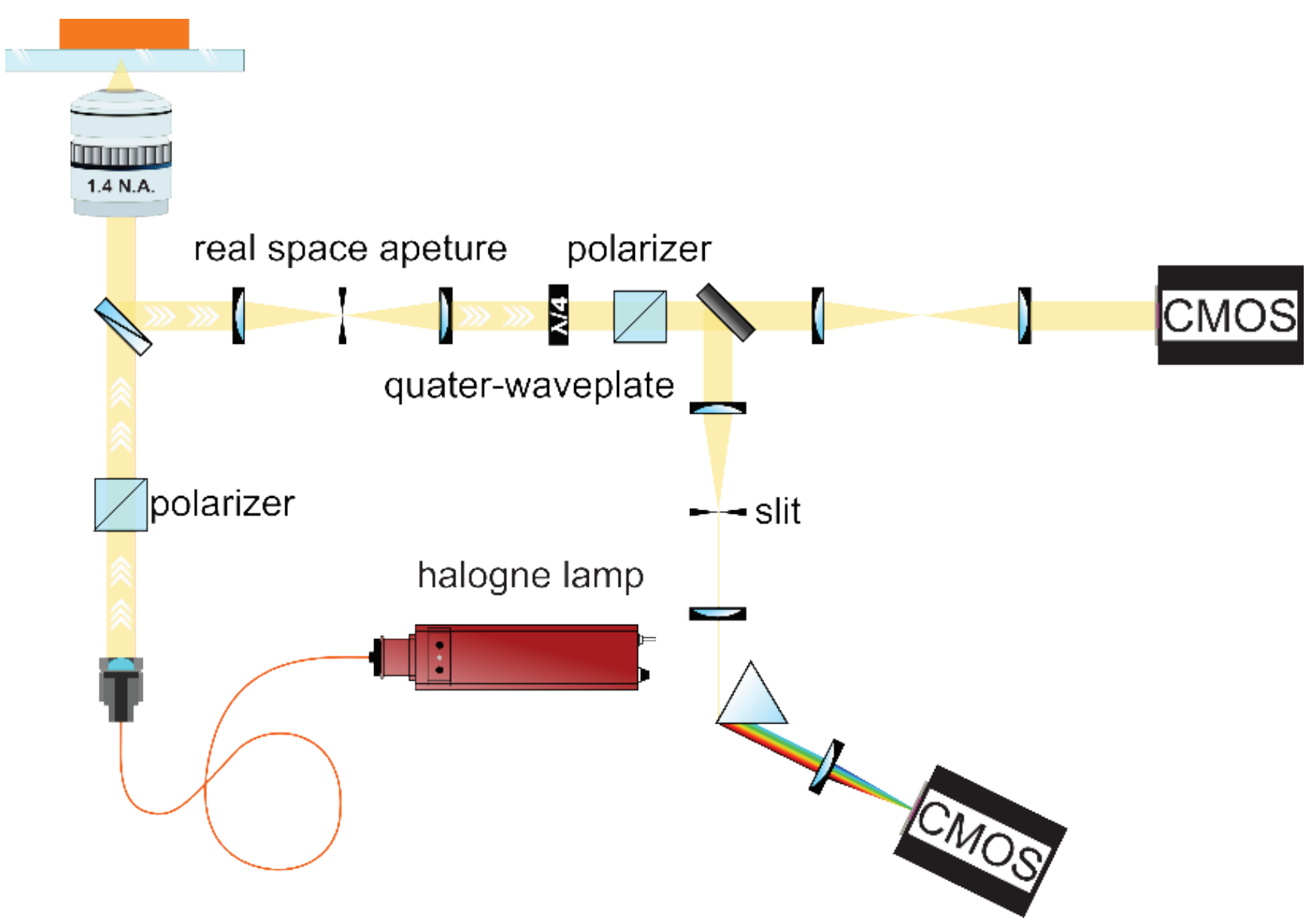

**Figure S2. Setup schematic for full Stokes vector imaging**. Stokes vector imaging is performed using a refractive microscope integrated with a polarization analyzing component placed at either the real-space image plane or the back-focal plane, enabling spatial- or momentum-resolved polarization mapping. The light source is a stabilized tungsten-halogen lamp (Thorlabs, SLS201L; 360-2600 nm). A Glan-Taylor polarizer (Thorlabs, GT15) was placed after the lamp to control the incident polarization. For angle-resolved spectral measurements, a real-space aperture (Thorlabs, P200K; pinhole diameter $200 \pm 6$ µm) was placed in the imaging plane to create a quasi-point source. The polarization analyzer consists of a rotating quarter-wave plate (QWP) followed by a fixed linear polarizer, following the procedure in Schaefer *et al.*[1] The QWP introduces a relative phase shift between orthogonal polarization components of the reflected field. For each measurement, we acquired 18 images at QWP fast axis rotating from 0° to 170° in 10° steps. The QWP was mounted on a motorized rotation stage synchronized with a cooled CMOS camera (Thorlabs, CC505MU), which recorded one image at each analyzer angle.

## 2. Stokes parameters reconstruction of outcoupled waveguided light

Because $NbOI_2$ has a very high refractive index (> 3.5), the polarization state of the waveguided light is modified upon refraction at the waveguide edge. Indeed, edge refraction strongly favors the p-polarized component due to near-Brewster transmission, leading to a systematic distortion of the experimentally measured Stokes parameters. To recover the state of the waveguided light at the edge *prior to* refraction, we apply a Mueller-matrix correction that accounts for the polarization-dependent transmission at the interface. The intrinsic Stokes vector $\boldsymbol{S}_{\text{initial}}$ is related to the measured Stokes vector $\boldsymbol{S}_{\text{meas}}$ by $\boldsymbol{S}_{\text{meas}} = \boldsymbol{M} \cdot \boldsymbol{S}_{\text{initial}}$, where $\boldsymbol{M}$ is the Muller matrix describing refraction at the $NbOI_2$/air interface with dielectric constant and incident angle as $(\varepsilon_x, \varepsilon_y, \varepsilon_z; \theta)$ = (5.0, 13.13, 6.15, 80º) for the 1.3 eV FW light, and $(\varepsilon_x, \varepsilon_y, \varepsilon_z; \theta)$ = (5.0, 13.95, 8.336, 80º) for the 2.6 eV SH light. As shown in Fig. S3, the linear polarization ($S_1 > 0.7$, Fig. S3a-b) is reduced to a more physically reasonable level after this correction ($S_1 \sim 0.3$, Fig. S3c-d), while the helicity component $S_3$ preserves its spatial phase and symmetry and is only moderately enhanced (~20%). This demonstrates that the conversion corrects the polarization imbalance introduced by refraction rather than generating artificial SOI. Consistently, the second-harmonic signal in Fig. S4a-b, which are predominantly s-polarized, exhibits minimal change in its Stokes parameters after conversion, confirming that the correction primarily compensates the refraction-induced amplification of the p-polarized component. Note that this conversion is only valid for the outcoupled light at the upper edge of the waveguide – the correction factor is meaningless for other locations in the image.

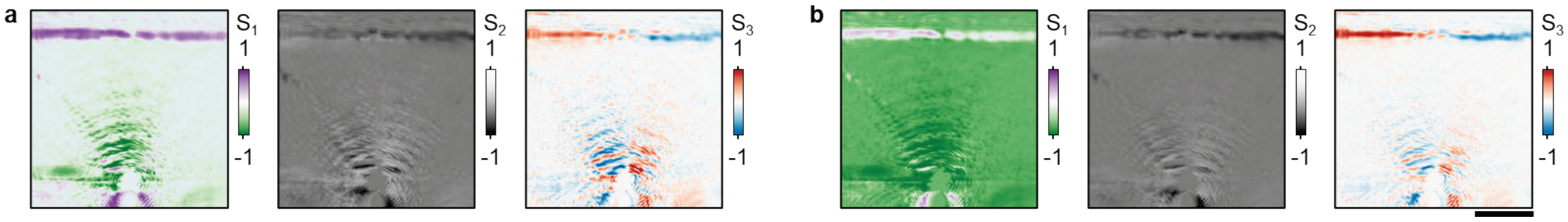

**Figure S3.** The raw spatially resolved Stokes vector ($S_1$, $S_2$, $S_3$) of the waveguided fundamental light are shown in (a), and the reconstructed Stokes vector images of waveguided fundamental light are shown in (b). The scale bar is 10 μm.

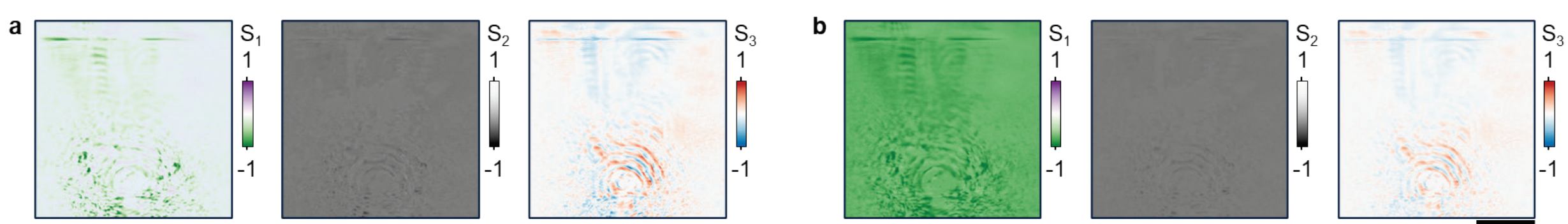

**Figure S4.** The raw spatially resolved Stokes vector ($S_1$, $S_2$, $S_3$) of the waveguided second-harmonic light are shown in (a), and the reconstructed Stokes vector images of waveguided second-harmonic light are shown in (b). The scale bar is 10 μm.

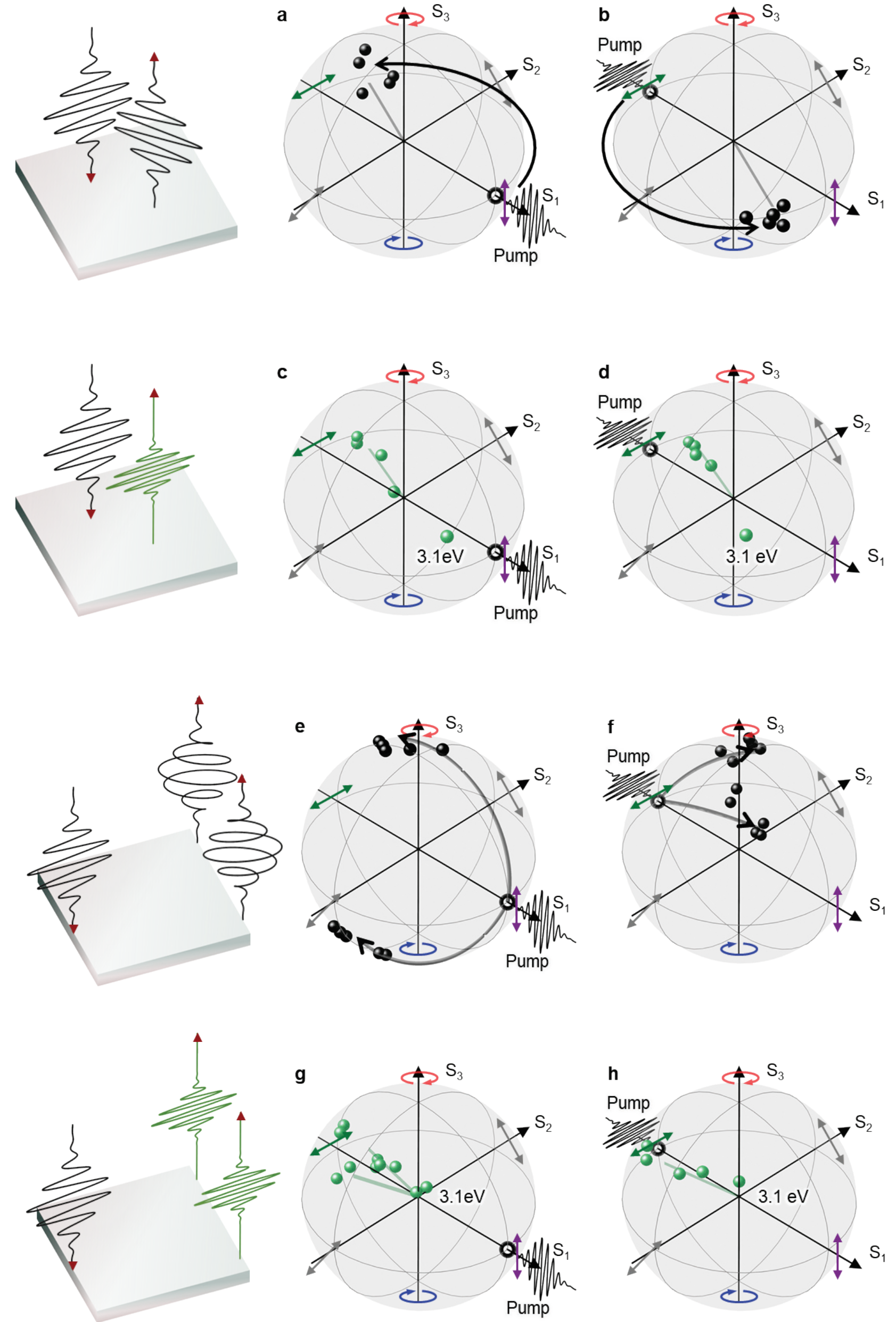

**Figure S5. $NbOI_2$ waveguide as a half waveplate and spin splitter**. Stoke vectors of fundamental and second harmonic light from $NbOI_2$ are shown on the Poincaré sphere. Pump energies range from 1.20 eV to 1.54 eV. The linear polarization orientations of the pump laser are indicated by black open circles. Fundamental (FW) and second harmonic (SH) light are indicated by black balls and green balls, respectively. Schematic insets show the experimental configuration, using either normally incident excitation and detection, or waveguide excitation. The FW and SH light are represented by black and green pulses. The measurements show that the $NbOI_2$ crystal acts as a half waveplate for FW light for normal incidence in panels a-b, as a half waveplate for SH light for normal incidence in panels c-d. The SH light at 3.08 eV near the band edge starts to depolarize, matching the observation in previous reports of $NbOI_2$ as SH generation crystals.[2] Under waveguide excitation, the $NbOI_2$ crystal acts as spin-splitter for FW light in panels e-f, and as a linear polarizer for SH light in panels g-h.

## 3. Simulating spin-orbit interactions in multimode cavities via a microscopic light-matter coupling model

### 3.1 Microscopic anisotropic light-matter coupling Hamiltonian

To construct the microscopic light-matter coupling Hamiltonian (MLMC), we consider a slab of dielectric materials with thickness of $L_z$ along the $\vec{z}$ direction and extended in the $xy$-plane with a refractive index higher than its surrounding medium. Solutions of Maxwell equations for this system are transverse electric field mode (TE), where vector of electric field is perpendicular to the plane of propagation, and transverse magnetic field mode (TM) where vector of magnetic field is perpendicular to the plane of propagation (plane of propagation is the plane made by the vectors of incident and refracted (or reflected) rays). Assuming $L_z$ to be much smaller than slab dimensions along $\vec{x}$ and $\vec{y}$ directions, the electric field vectors for a right-travelling plane wave within this slab of dielectric are given by:[3]

$$\vec{E}_{TE}\big(\vec{k}(q),t\big) = g\sin\big(k_{q,z}r_z\big)\, e^{i(\vec{k}_{\parallel}\cdot\vec{R}-\omega_{q,k}t)}\{\vec{e}_{\parallel}\times\vec{z}\},$$

$$\vec{E}_{TM}\big(\vec{k}(q),t\big) = g\left[\frac{c|\vec{k}_{\parallel}|}{\eta\omega_{q,k}}\cos\big(k_{q,z}r_z\big)\vec{z} - \frac{ick_z}{\eta\omega_{q,k}}\sin\big(k_{q,z}r_z\big)\,\vec{e}_{\parallel}\right]e^{i(\vec{k}_{\parallel}\cdot\vec{R}-\omega_{q,k}t)}$$

where $c$ is the speed of light in free space, $\eta$ is dielectric background refractive index, $g$ is a real scalar constant, and $\vec{R} = r_x\vec{x} + r_y\vec{y} + r_z\vec{z}$. Here, $k_{q,z} = \frac{q\pi}{L_z}$ ($q$ is a positive integer) is the confined field wavevector along quantization direction ($z$ direction). In this setup, $\vec{k}(q) = \vec{k}_{\parallel} + k_{q,z}\vec{z} = k_x\vec{x} + k_y\vec{y} + k_{q,z}\vec{z}$ is the confined electromagnetic field wavevector with frequency of $\omega_{q,k} = \frac{c}{\eta}|\vec{k}(q)| = \frac{c}{\eta}\sqrt{\left|\vec{k}_{\parallel}\right|^2 + k_{q,z}^2}$. Here $\vec{k}_{\parallel}$ is the in-plane ($xy$-plane) wavevector components with the unit vector of $\vec{e}_{\parallel} = \vec{k}_{\parallel}/\left|\vec{k}_{\parallel}\right|$. Note that we treat this dielectric cavity to be lossless for simplicity, although this assumption does not change any of the conclusions of our work and adding loss is straightforward. In this work, we extract parameters directly from our experimental results of a stacked ferroelectric van der Waals materials, namely $NbOI_2$, which forms a dielectric (open) cavity. In our prior work[4] we demonstrate that the light-matter coupling between photons and thick dielectric materials can be mapped to a set of uncoupled light-matter systems composed of an single effective delocalized layer (which is a 2D rectangular lattice) and a cavity mode. Each of the layers of the material is described with $N_x$ sites along $\vec{x}$ direction with lattice constant of $a_x$, and $N_y$ site along $\vec{y}$ direction with lattice constant of $a_y$. Using a periodic boundary condition in the $\vec{x}$ and $\vec{y}$ direction, we write $k_x = \frac{2\pi m_x}{N_x a_x}$ and $k_y = \frac{2\pi m_y}{N_y a_y}$ for $m_x \in \left[-\frac{N_x}{2}, -\frac{N_x}{2}+1, \dots, -\frac{N_x}{2}-1\right]$ and $m_y \in \left[-\frac{N_y}{2}, -\frac{N_y}{2}+1, \dots, -\frac{N_y}{2}-1\right]$ for even $N_x$ and $N_y$. Since in $NbOI_2$, in-plane lattice vectors ($b$ and $c$) are perpendicular to each other, for the rest of this section we label, $k_x \rightarrow k_b$ and $k_y \rightarrow k_c$, thus $\vec{k}_\parallel = k_b\, \vec{e}_b + k_c\, \vec{e}_c$. We also use $r_x = n_b|b|$ and $r_y = n_c|c|$ and write $R_n = n_b|b|\vec{e}_b + n_c|c|\vec{e}_c + n_z a_z \vec{z}$ where $\vec{e}_b$ and $\vec{e}_c$ are the unit vectors, and $\vec{e}_z \parallel (\vec{e}_b \times \vec{e}_c)$. Here $n_b \in \left[-\frac{N_x}{2}, -\frac{N_x}{2}+1, \dots, -\frac{N_x}{2}-1\right]$ and $n_c \in \left[-\frac{N_y}{2}, -\frac{N_y}{2}+1, \dots, -\frac{N_y}{2}-1\right]$.

Quantizing electric field within the slab of the dielectric material is given by transforming $e^{i(\vec{k}_\parallel \cdot \vec{R}_n - \omega_{q,k} t)} \rightarrow \hat{a}_{q,\boldsymbol{k}} e^{i\vec{k}_\parallel \cdot \vec{R}_n}$ in $\vec{E}_{TM}$ and $e^{i(\vec{k}_\parallel \cdot \vec{R}_n - \omega_{q,k} t)} \rightarrow \hat{b}_{q,\boldsymbol{k}} e^{i\vec{k}_\parallel \cdot \vec{R}_n}$ in $\vec{E}_{TM}$. Here, $\hat{a}^\dagger_{q,\boldsymbol{k}}$ create a TE photon with transition frequency of $\omega_{q,k}$ and $\hat{b}^\dagger_{q,\boldsymbol{k}}$ create an TM photon mode with the same transition frequency of $\omega_{q,k}$. Using the quantized description of the field, we can decompose the TE and TM modes in $\vec{b}$ and $\vec{c}$ direction as

$$\vec{E}_b = g_q\left(k_c \hat{a}_{q,\boldsymbol{k}} - i\frac{ck_z}{\eta\omega_{q,k}} k_b \hat{b}_{q,\boldsymbol{k}}\right)\frac{1}{|\vec{k}_\parallel|} e^{i\vec{k}_\parallel \cdot \vec{R}_n} + h.c.,$$

$$\vec{E}_c = g_q\left(-k_b \hat{a}_{q,\boldsymbol{k}} - i\frac{ck_z}{\eta\omega_{q,k}} k_c \hat{b}_{q,\boldsymbol{k}}\right)\frac{1}{|\vec{k}_\parallel|} e^{i\vec{k}_\parallel \cdot \vec{R}_n} + h.c.$$

where we ignored the electric field along $\vec{z}$ direction ($\vec{E}_z$ from TM mode) since we have already incorporated the material transition dipoles in the $\vec{z}$ direction within the background refractive index. Considering this quantized picture, the bare photonic Hamiltonian of $q$'th quantization mode is given by

$$H_{ph}(q, \boldsymbol{k}) = \sum_{\boldsymbol{k}} \omega_{q,\boldsymbol{k}}\left(\hat{a}^\dagger_{q,\boldsymbol{k}} \hat{a}_{q,\boldsymbol{k}} + \hat{b}^\dagger_{q,\boldsymbol{k}} \hat{b}_{q,\boldsymbol{k}}\right)$$

where we defined $\boldsymbol{k} = (k_b, k_c)$. The bare excitonic Hamiltonian coupled to TE and TM modes is a 2D tight-binding model with $N_d$ excitons per unit cell (site), which is given by

$$H_{ex} = \sum_{\ell}^{N_d}\left(\epsilon_\ell \sum_{\boldsymbol{n}} \hat{X}^\dagger_{\ell,\boldsymbol{n}} \hat{X}_{\ell,\boldsymbol{n}} - \tau \sum_{\boldsymbol{n}}\left(\hat{X}^\dagger_{\ell,\boldsymbol{n}} \hat{X}_{\ell,\boldsymbol{n}'} + h.c.\right)\right)$$

where $\boldsymbol{n} = (n_b, n_c)$, $\tau = 0.0125$ eV is the hopping parameter, and $\epsilon_\ell$ is the $\ell$'th exciton on-site energy. Here $\hat{X}^\dagger_{\ell,\boldsymbol{n}}$ creates $\ell$'th exciton at $\boldsymbol{n}$'th site and we only consider nearest neighbor in-plane hopping such that $\hat{X}^\dagger_{\ell,\boldsymbol{n}} \hat{X}_{\ell,\boldsymbol{n}'}$ only survives when $n'_b = n_b \pm 1$ or $n'_c = n_c \pm 1$. This tight binding model is diagonalizable in momentum space ($\boldsymbol{k}$-space), by Fourier transforming the excitonic operators using $\hat{X}_{\ell,\boldsymbol{k}} = \frac{1}{\sqrt{N}} \sum_{\boldsymbol{n}} \hat{X}_{\ell,\boldsymbol{n}} e^{i\vec{k}_\parallel \cdot \vec{R}_n}$. The block diagonal form of $H_{ex}$ is given by

$$H_{ex}(\boldsymbol{k}) = \sum_{\ell,\boldsymbol{k}} \epsilon_{\ell,\boldsymbol{k}}\, \hat{X}^\dagger_{\ell,\boldsymbol{k}} \hat{X}_{\ell,\boldsymbol{k}}$$

where $\epsilon_{\ell,\boldsymbol{k}} = \left(\epsilon_\ell - 2\tau(\cos(k_b b) + \cos(k_c c))\right)$. The light-matter coupling of $q$ 'th photon quantization mode with excitons is given by

$$H_{ex-ph}(n,q,\boldsymbol{k}) = -\sum_\ell \vec{\mu}_\ell \cdot \vec{E} = \frac{g\sqrt{\omega_{q,0}}}{\sqrt{N}} \sum_{\boldsymbol{n}} \sum_{\boldsymbol{k}} \frac{1}{|\vec{k}_\parallel|} \Bigg( \sum_\ell \mu_\ell^b \left[ \hat{X}_{\ell,\boldsymbol{n}} \left( \sqrt{\frac{\omega_{q,\boldsymbol{k}}}{\omega_{q,0}}} k_c \hat{a}_{q,\boldsymbol{k}}^\dagger - i\sqrt{\frac{\omega_{q,0}}{\omega_{q,\boldsymbol{k}}}} k_b \hat{b}_{q,\boldsymbol{k}}^\dagger \right) e^{i\vec{k}_\parallel \cdot \vec{R}_n} \right] - \sum_\ell \mu_\ell^c \left[ \hat{X}_{\ell,\boldsymbol{n}} \left( \sqrt{\frac{\omega_{q,\boldsymbol{k}}}{\omega_{q,0}}} k_b \hat{a}_{q,\boldsymbol{k}}^\dagger + i\sqrt{\frac{\omega_{q,0}}{\omega_{q,\boldsymbol{k}}}} k_c \hat{b}_{q,\boldsymbol{k}}^\dagger \right) e^{i\vec{k}_\parallel \cdot \vec{R}_n} \right] + h.c. \Bigg).$$

Here $(\vec{\mu}_\ell = \mu_\ell^b\, \vec{b} + \mu_\ell^c\, \vec{c})$, $N = N_x N_y$ and $\omega_{q,0} = \omega_{q,\boldsymbol{k}=0} = \frac{ck_{q,z}}{\eta}$. Here, $H_{ex-ph}(n,\boldsymbol{k})$ is also block-diagonalizable in the momentum space, given by

$$H_{ex-ph}(q,\boldsymbol{k}) = g_{q,0} \sum_{\boldsymbol{k}} \frac{1}{|\vec{k}_\parallel|} \sum_\ell \Bigg( \mu_\ell^b \left[ \hat{X}_{\ell,\boldsymbol{k}} \left( \sqrt{\frac{\omega_{q,\boldsymbol{k}}}{\omega_0}} k_c \hat{a}_{q,\boldsymbol{k}}^\dagger - i\sqrt{\frac{\omega_0}{\omega_{\boldsymbol{k}}}} k_b \hat{b}_{q,\boldsymbol{k}}^\dagger \right) \right] - \mu_\ell^c \left[ \hat{X}_{\ell,\boldsymbol{k}} \left( \sqrt{\frac{\omega_{q,\boldsymbol{k}}}{\omega_{q,0}}} k_b \hat{a}_{q,\boldsymbol{k}}^\dagger + i\sqrt{\frac{\omega_{q,0}}{\omega_{q,\boldsymbol{k}}}} k_c \hat{b}_{q,\boldsymbol{k}}^\dagger \right) \right] + h.c. \Bigg)$$

where $g_{q,0} = g\sqrt{\omega_{q,0}}$ is the light-matter coupling constant. The full MLMC Hamiltonian then is given by

$$\hat{H}_{LM}(q,\boldsymbol{k}) = \hat{H}_{ex}(\boldsymbol{k}) + \hat{H}_{ph}(q,\boldsymbol{k}) + \hat{H}_{ex-ph}(q,\boldsymbol{k}).$$

This Hamiltonian is block-diagonal in $\boldsymbol{k}$-space and for each photon quantization number $q$, where each block with matrix dimension of $(N_d+2)\times(N_d+2)$ is given by

$$\hat{H}_{LM}(q,\boldsymbol{k}) = \begin{pmatrix} H_{ex}(\boldsymbol{k}) & H_I(q,\boldsymbol{k}) \\ H_I^*(q,\boldsymbol{k}) & H_{ph}(q,\boldsymbol{k}) \end{pmatrix} = \begin{pmatrix} \epsilon_{1,\boldsymbol{k}} & 0 & \cdots & 0 & h_{1,\boldsymbol{k}}(q) & g_{1,\boldsymbol{k}}(q) \\ 0 & \epsilon_{2,\boldsymbol{k}} & & 0 & h_{2,\boldsymbol{k}}(q) & g_{2,\boldsymbol{k}}(q) \\ \vdots & & \ddots & & & \vdots \\ 0 & 0 & & \epsilon_{N_d,\boldsymbol{k}} & h_{N_d,\boldsymbol{k}}(q) & g_{N_d,\boldsymbol{k}}(q) \\ h_{1,\boldsymbol{k}}^*(q) & h_{2,\boldsymbol{k}}^*(q) & \cdots & h_{N_d,\boldsymbol{k}}^*(q) & \omega_{q,\boldsymbol{k}} & 0 \\ g_{1,\boldsymbol{k}}^*(q) & g_{2,\boldsymbol{k}}^*(q) & & h_{N_d,\boldsymbol{k}}^*(q) & 0 & \omega_{q,\boldsymbol{k}} \end{pmatrix}.$$

Here, $H_I(q,\boldsymbol{k})$ is the anisotropic exciton-photon interaction terms and is given by

$$H_I(q,\boldsymbol{k}) = \begin{pmatrix} h_{1,\boldsymbol{k}}(q) & g_{1,\boldsymbol{k}}(q) \\ h_{2,\boldsymbol{k}}(q) & g_{2,\boldsymbol{k}}(q) \\ \vdots & \vdots \\ h_{N_d,\boldsymbol{k}}(q) & g_{N_d,\boldsymbol{k}}(q) \end{pmatrix}.$$

Here $h_{\ell,\boldsymbol{k}}(q) = \frac{g_{q,0}}{|\vec{k}_\parallel|}\sqrt{\frac{\omega_{q,k}}{\omega_{q,0}}}\left[\mu_\ell^b k_c - \mu_\ell^c k_b\right]$, which is the excitonic coupling to the TE modes, and $g_{\ell,\boldsymbol{k}}(q) = \frac{-ig_{q,0}}{|\vec{k}_\parallel|}\sqrt{\frac{\omega_{q,0}}{\omega_{q,k}}}\left[\mu_\ell^b k_b + \mu_\ell^c k_c\right]$ is the excitons coupling with the TM modes.

In our numerical simulations we use $NbOI_2$ crystal structure with crystal lattice constants of $b = 3.94\text{Å}$ and $c = 7.51\text{Å}$. Additionally, we extract the background refractive index, excitons onsite (transition) energy and the transition dipoles ($\vec{\mu}_\ell = \mu_\ell^b\,\vec{b} + \mu_\ell^c\,\vec{c}$) from experimentally measured dielectric function by fitting to the following theoretical Lorentz oscillator model

$$\varepsilon(\omega) = \varepsilon_\infty + \sum_l \frac{\epsilon_l |\vec{\mu}_l|^2}{\epsilon_l^2 - \omega^2}$$

here ω is optical frequency (assuming $\hbar = 1$ unit of ω is energy), $\ell$ runs the summation on transition dipoles from ground to excited states for $\epsilon_\ell$ being complex transition resonance energies that encode finite lifetime ($\epsilon_\ell = \Re(\epsilon_\ell) + i\Im(\epsilon_\ell)$), $\vec{\mu}_\ell$ are exciton transition dipole moments and $\varepsilon_\infty = \eta^2$. We numerically fit this model to experimental dielectric function, from which we can properly extract a set of $\epsilon_\ell$ and $\vec{\mu}_\ell$. Fig. S6 presents curve fitting of dielectric function results.

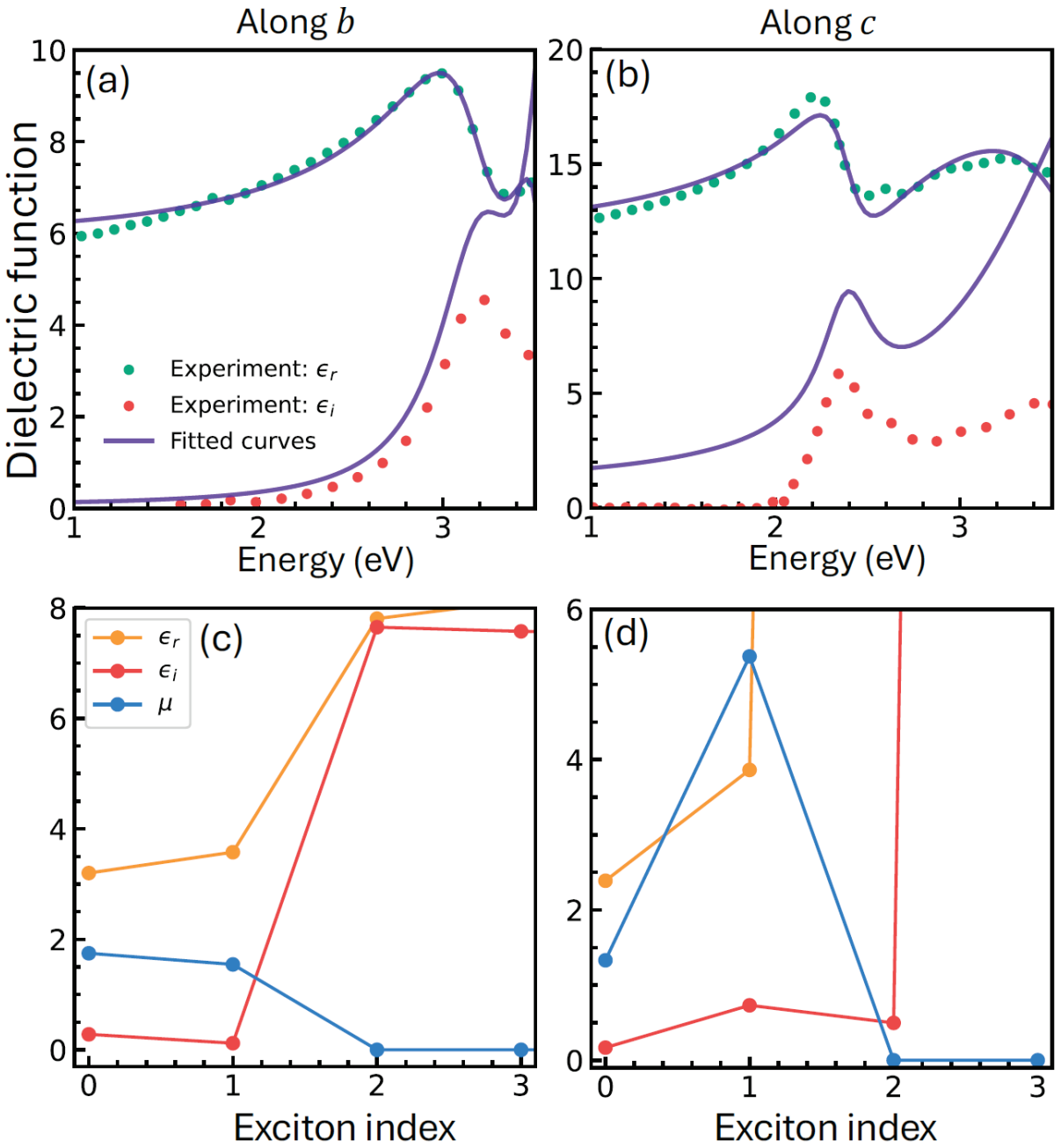

**Figure S6. Transition dipoles on-site energy and transition dipole moments.** (a) and (b) are the experimental dielectric function (dots) and fitted curve. (c) and (d) are the parameters used to obtain the fitted curve for each case.

The parameters producing best fit for the dielectric function and energy dispersion is obtained for $\eta \approx \sqrt{4.5}$ for two independent excitons along lattice vector $b$ and two along lattice vector $c$, which are given in Table 1.

**Table 1**. Excitons on-site energy and transition dipole moment obtained from curve fitting the experimental dielectric function

| | Exciton 1 | Exciton 2 | Exciton 3 | Exciton 4 |
|---|---|---|---|---|
| direction | $b$ | $b$ | $c$ | $c$ |
| $\Re(\epsilon_\ell)$(eV) | 3.2 | 3.58 | 2.39 | 3.86 |
| $\lvert\vec{\mu}\rvert$ | 1.75 | 1.55 | 1.33 | 5.37 |

To numerically calculate the energy dispersion of MLMC model presented in the main text, we used five photon quantization modes, i.e. $q \in \{1,2,3,4,5\}$ with different values of $g_{q,0}$ and different thicknesses of dielectric slab, $L_{q,z}$. The optimized $g_{q,0}$ and $L_{q,z}$ to give dispersion with best overlap with experiment are given in Table 2.

**Table 2**. Optimized parameters for the numerically calculated energy dispersion

| $q$ | 1 | 2 | 3 | 4 | 5 |
|---|---|---|---|---|---|
| $\omega_{q,0}$(eV) | 0.468 | 0.59 | 0.578 | 0.627 | 0.636 |
| $L_{q,z}$($\mu$m) | 0.625 | 0.495 | 0.506 | 0.466 | 0.46 |
| $g_{q,0}$(eV) | 0.113 | 0.162 | 0.158 | 0.184 | 0.184 |

**3.2 Quantum dynamics**

The time-dependent wavefunction, evolved with the MLMC Hamiltonian is given by

$$|\psi(t)\rangle = \sum_{\boldsymbol{k},\ell}\left(\sum_{q}\left(c_{TE}(q,\boldsymbol{k},t)\hat{a}^{\dagger}_{q,\boldsymbol{k}} + c_{TM}(q,\boldsymbol{k},t)\hat{b}^{\dagger}_{q,\boldsymbol{k}}\right) + d_{\ell,\boldsymbol{k}}(t)\hat{X}^{\dagger}_{\boldsymbol{\ell},\boldsymbol{k}}\right)|\bar{0}\rangle$$
$$= \sum_{\boldsymbol{k},\ell}\left(\sum_{q}\left(c_{TE}(q,\boldsymbol{k},t)|TE,q,\boldsymbol{k}\rangle + c_{TM}(q,\boldsymbol{k},t)|TM,q,\boldsymbol{k}\rangle\right) + d_{\ell,\boldsymbol{k}}(t)|\ell,\boldsymbol{k}\rangle\right)$$

where we defined $|TE,q,\boldsymbol{k}\rangle = \hat{a}^{\dagger}_{q,\boldsymbol{k}}|\bar{0}\rangle$, $|TM,q,\boldsymbol{k}\rangle = \hat{b}^{\dagger}_{q,\boldsymbol{k}}|\bar{0}\rangle$, and $|\ell,\boldsymbol{k}\rangle = \hat{X}^{\dagger}_{\boldsymbol{\ell},\boldsymbol{k}}|\bar{0}\rangle$ with $|\bar{0}\rangle$ as the vacuum state of entire light-matter hybrid system corresponding to no excitons or photons. The time-evolution of $|\psi_{q,\boldsymbol{k}}(0)\rangle$ is given by

$$|\psi_{q,\boldsymbol{k}}(t)\rangle = e^{-i\hat{H}(q,\boldsymbol{k})t}|\psi_{q,\boldsymbol{k}}(0)\rangle.$$

To generate the initial state, we use a linear superposition of two lowest-polariton states which is written as

$$|\psi_{\boldsymbol{k}}(0)\rangle = \sum_{k\in E_{int}(\boldsymbol{k})} c_{k,0}\hat{P}^{\dagger}_{k,0}|\bar{0}\rangle + \sum_{k\in E_{int}(\boldsymbol{k})} c_{k,1}\hat{P}^{\dagger}_{k,1}|\bar{0}\rangle,$$

where $E_{int}(k)$ denotes a subspace of initial states with $\vec{k}_{\parallel}$ drawn from a finite set of $\boldsymbol{k}$-points where the energy of lowest polariton band ($E_0(k)$) and second to lowest polariton band ($E_1(k)$) lie within $E_{init} - \frac{\delta E}{2} < E_{0,1}(k) < E_{init} + \frac{\delta E}{2}$. Here, $\hat{P}^{\dagger}_{k,j}$ generate a polariton of wavevectors $\boldsymbol{k}$ from band $j$, where the polariton bands obtained from diagonalizing $\widehat{H}_{LM}(q,\boldsymbol{k})$. The coefficients $c_{k,0}$ and $c_{k,1}$ are obtained by fitting the photonic real space wavefunction to a Gaussian function written as

$$f(r_x, r_y) \propto \exp\left(\frac{|\vec{R}_{\parallel}-\vec{r}_0|^2}{2\sigma^2}\right),$$

where $\vec{R}_{\parallel} = r_x\vec{x} + r_y\vec{y}$ is the in-plane real space coordinate, $\vec{r}_0 = \frac{N_x b}{2}\vec{x} + \frac{N_y c}{2}\vec{y}$ is the center of the Gaussian wave packet with variance of $\sigma^2$.

The photonic subspace of exciton-photon state $|\psi_{q=1,\boldsymbol{k}}(t)\rangle$ (and in general $|\psi_{q,\boldsymbol{k}}(t)\rangle$) is in a linearly polarized basis of TE and TM polarization modes. To find the spin of light for such system using the Stokes parameters we can define circularly polarized states written as

$$|R,q,\boldsymbol{k}\rangle = \frac{1}{\sqrt{2}}(|TE,q,\boldsymbol{k}\rangle + i|TM,q,\boldsymbol{k}\rangle)$$

$$|L,q,\boldsymbol{k}\rangle = \frac{1}{\sqrt{2}}(|TE,q,\boldsymbol{k}\rangle - i|TM,q,\boldsymbol{k}\rangle),$$

where $|R,q,\boldsymbol{k}\rangle$ and $|L,q,\boldsymbol{k}\rangle$ are right and left photonic circularly polarized states, respectively. Accordingly, for $q=1$ we can define Identity ($\hat{I}$) and Pauli ($\hat{\sigma}_x$, $\hat{\sigma}_y$, and $\hat{\sigma}_z$) matrices using circularly polarized states by

$$\hat{I}(\boldsymbol{k}) = |R,1,\boldsymbol{k}\rangle\langle R,1,\boldsymbol{k}| + |L,1,\boldsymbol{k}\rangle\langle L,1,\boldsymbol{k}$$

$$\hat{\sigma}_x(\boldsymbol{k}) = |L,1,\boldsymbol{k}\rangle\langle R,1,\boldsymbol{k}| + |R,1,\boldsymbol{k}\rangle\langle L,1,\boldsymbol{k}|$$

$$\hat{\sigma}_y(\boldsymbol{k}) = i|L,1,\boldsymbol{k}\rangle\langle R,1,\boldsymbol{k}| - i|R,1,\boldsymbol{k}\rangle\langle L,1,\boldsymbol{k}|$$

$$\hat{\sigma}_z(\boldsymbol{k}) = |R,1,\boldsymbol{k}\rangle\langle R,1,\boldsymbol{k}| - |L,1,\boldsymbol{k}\rangle\langle L,1,\boldsymbol{k}|$$

Therefore, the Stokes parameters at each $\boldsymbol{k}$-point and time $t$ are given by

$$S_0(\boldsymbol{k},t) = \langle\hat{I}(\boldsymbol{k})\rangle,\ S_1(\boldsymbol{k},t) = \langle\hat{\sigma}_x(\boldsymbol{k})\rangle,\ S_2(\boldsymbol{k},t) = \langle\hat{\sigma}_y(\boldsymbol{k})\rangle,\ S_3(\boldsymbol{k},t) = \langle\hat{\sigma}_z(\boldsymbol{k})\rangle$$

where $\langle\cdot\rangle$ stands for the expectation value using $|\psi_{q=1,\boldsymbol{k}}(t)\rangle$. Within this formalism, at time $t$ and at each $\boldsymbol{k}$-point Stoke's parameters satisfy $S_0^2(\boldsymbol{k},t) = S_1^2(\boldsymbol{k},t) + S_2^2(\boldsymbol{k},t) + S_3^2(\boldsymbol{k},t)$.

The Hamiltonian of two coupled degenerate state can be written in form of a $2\times 2$ Hamiltonian using Pauli matrices[5–7] In this work we numerically verified that the band splitting of TE and TM modes can be approximately mapped to a $2\times 2$ Hamiltonian which is an effective Rashba-Dresselhaus Hamiltonian given by

$$\mathrm{H}_{\mathrm{RD}}(\boldsymbol{k}) = \begin{pmatrix} \omega_k + \delta_z & \alpha + \beta(k_b + ik_c)^2 \\ \alpha + \beta(k_b - ik_c)^2 & \omega_k - \delta_z \end{pmatrix} = \omega_k \mathbb{1}_{2\times 2} + \vec{G}\cdot\vec{\sigma}.$$

Here, $\alpha$ is the optical birefringence, $\beta$ controls the $\boldsymbol{k}$-dependent splitting of the TE and TM polarizations, and $\delta_z$ represents the Zeeman splitting. Here, $\omega_k = \omega_{q=1,k}$ is the degenerate energy of the left and the right circularly polarized modes, $\vec{\sigma} = (\hat{\sigma}_x, \hat{\sigma}_y, \hat{\sigma}_z)$ is the vector of the Pauli matrices and $\vec{G}$ is the effective magnetic field given by

$$\vec{G}(\boldsymbol{k}) = (\alpha + \beta(k_b^2 - k_c^2), -2\beta k_c k_c, \delta_z).$$

To find $\alpha$, $\beta$, and $\delta z$ we fitted the lowest polariton band of $\widehat{H}(q = 1, \boldsymbol{k})$. The best fitted parameters obtained for our system are $\alpha = 0.204$ eV, $\beta = 1.02$ meV(μm)$^2$, $\delta_z = 0$ eV. Using these fitted parameters, we calculated the Berry curvature and quantum geometry tensor of the waveguide modes. The Berry curvature ($B_z$) and quantum geometry tensor ($g_{ij}$ for $i, j \in \{b, c\}$) components of the eigenstate are expressed as:[7]

$$g_{bb}(\boldsymbol{k}) = \frac{\beta^2(k_c^2\left(\alpha - \left|\overrightarrow{k_\parallel}\right|^2\beta\right)^2 + \left|\overrightarrow{k_\parallel}\right|^2\delta_z^2)}{\left(\alpha^2 + 2(k_b^2 - k_c^2)\alpha\beta + \left|\overrightarrow{k_\parallel}\right|^4\beta^2 + \delta_z^2\right)^2}$$

$$g_{cc}(\boldsymbol{k}) = \frac{\beta^2(k_b^2\left(\alpha + \left|\overrightarrow{k_\parallel}\right|^2\beta\right)^2 + \left|\overrightarrow{k_\parallel}\right|^2\delta_z^2)}{\left(\alpha^2 + 2(k_b^2 - k_c^2)\alpha\beta + \left|\overrightarrow{k_\parallel}\right|^4\beta^2 + \delta_z^2\right)^2}$$

$$g_{bc}(\boldsymbol{k}) = \frac{\beta^2 k_b k_c(\alpha^2 - \left|\overrightarrow{k_\parallel}\right|^4\beta^2)}{\left(\alpha^2 + 2(k_b^2 - k_c^2)\alpha\beta + \left|\overrightarrow{k_\parallel}\right|^4\beta^2 + \delta_z^2\right)^2}$$

$$B_z(\boldsymbol{k}) = \frac{2\beta^2\left|\overrightarrow{k_\parallel}\right|^2\delta_z}{\left(\alpha^2 + 2(k_b^2 - k_c^2)\alpha\beta + \left|\overrightarrow{k_\parallel}\right|^4\beta^2 + \delta_z^2\right)^{3/2}} = 0$$

The simulation results of the quantum geometry tensor are shown in Fig. S7.

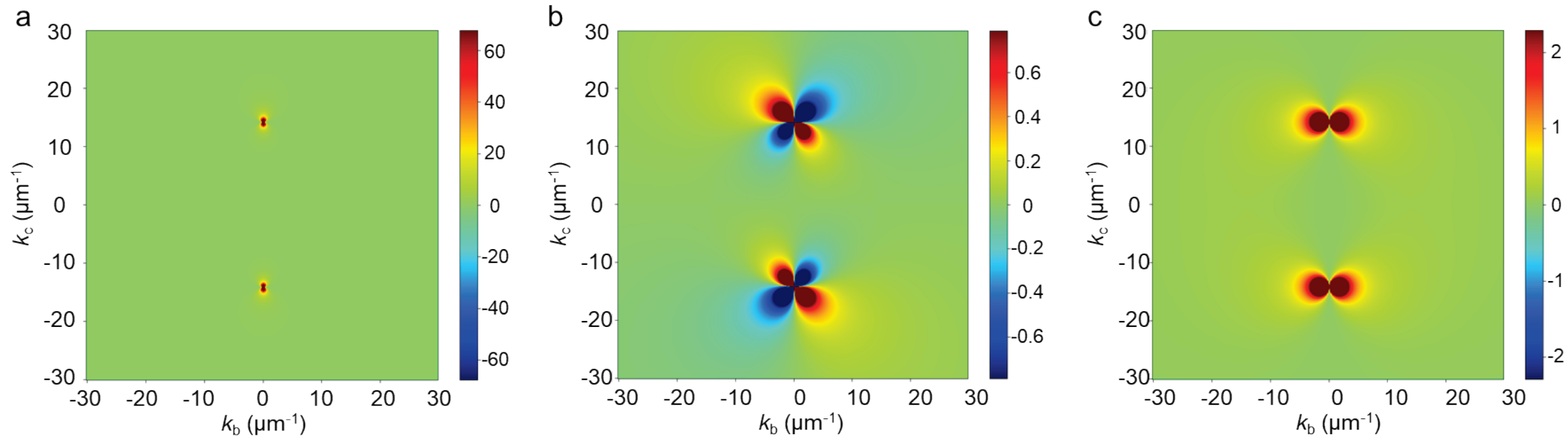

**Figure S7.** Different elements of quantum geometry tensors: $g_{bb}$ (a); $g_{bc}$ (b); and $g_{cc}$ (c).

To calculate the dynamics of Stokes parameters ($S_1$, $S_2$, and $S_3$) on the Poincaré sphere and the spin precession, we solved equation of motion of the form

$$\frac{\partial\vec{\sigma}(\boldsymbol{k})}{\partial t} = \vec{G}(\boldsymbol{k}) \times \vec{\sigma}(\boldsymbol{k})$$

In this equation the frequency of precession is controlled by the effective magnetic field.

### 3.3 Small deviations between theory and experiment

Quantitatively reproducing the dielectric response of a biaxial material remains challenging for the MLMC model. In $NbOI_2$, this difficulty likely arises because for tractability, the dielectric fit includes four exciton states and doesn't fully capture the multiple transitions at the semiconductor band edge (Fig. S6). As a result, the full dispersions in Extended Data Fig. 2 do not perfectly fit the transfer matrix simulations and experimental dispersions, and the light-matter coupling strength may be slightly over-estimated. These small deviations may explain the enlarged avoided-crossing gap and the narrower beam spatial separation in the calculations compared to experiments in Fig. 2b. Figure S8 below shows the calculated spin-resolved spatial propagation for three energies. The experimental result at 1.3 eV shows a closer match to the calculated results obtained between 0.9 – 1.1 eV due to these small dielectric function deviations.

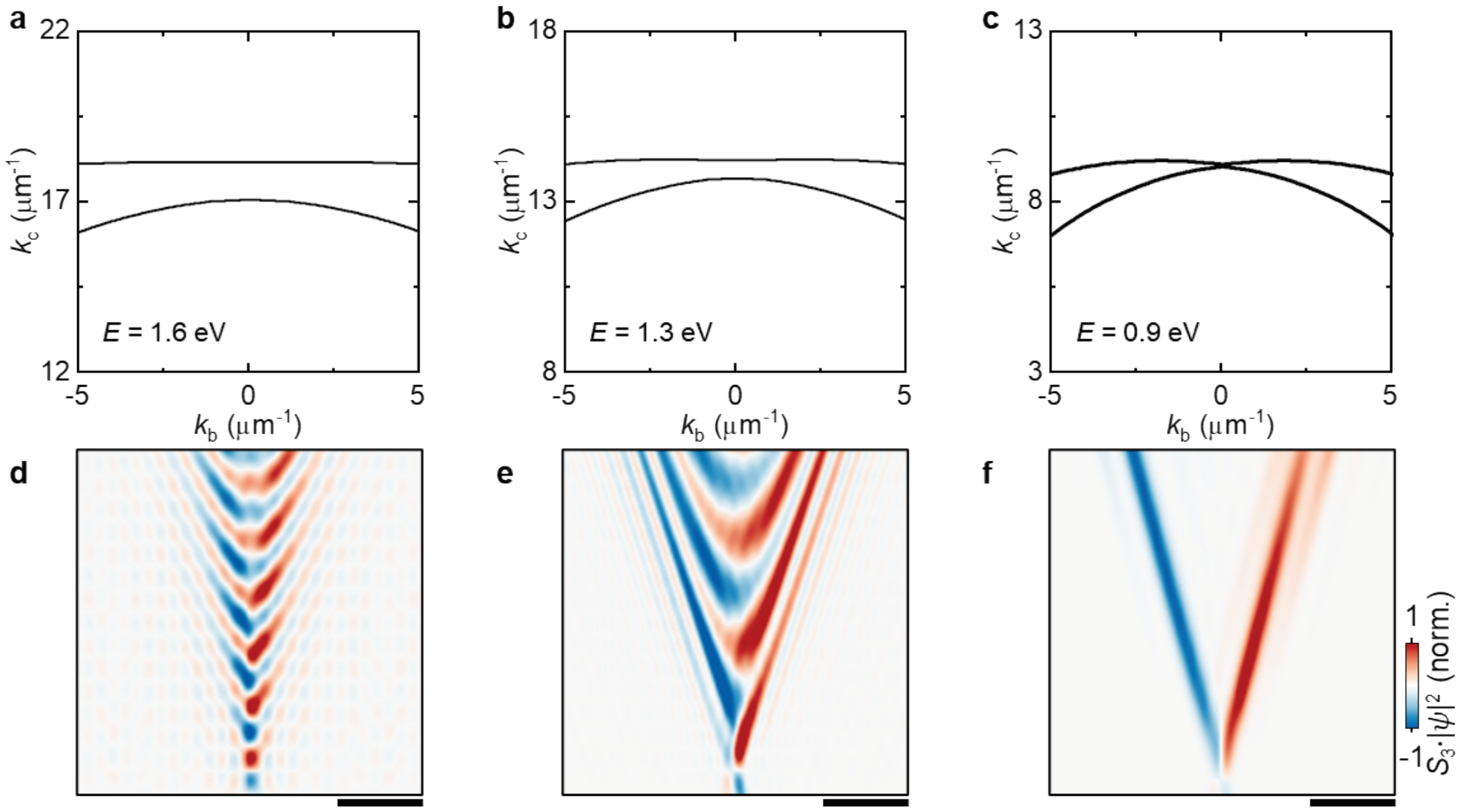

**Figure S8. Calculated spin-resolved propagation for different energies and crossings**. a-c, Isofrequency contours of the waveguided modes at 1.6, 1.3, and 0.9 eV. The mode structure evolves from a DP to an avoided crossing at higher energies. d-f, simulated $S_3$ spin polarization propagation for excitation at the (avoided) crossing for each energy.